\documentclass[11pt,hyper]{JHEP3}
\usepackage[ansinew]{inputenc}
\usepackage{amsfonts,amssymb,amsmath,amsthm,cite}
\usepackage{graphicx}
\usepackage{epsfig}
\usepackage{rotating}

\newcommand{\rhotilde}{\tilde{\rho}}
\newcommand{\mtilde}{\tilde{m}}
\newcommand{\Mtilde}{\tilde{M}}

\newcommand{\ctilde}{\tilde{c}}
\newcommand{\Btilde}{\tilde{B}}
\newcommand{\rhoIR}{\rho_{IR}}
\newcommand{\omIR}{\om_{IR}}
\newcommand{\LIR}{L_{IR}}
\newcommand{\rIR}{r_{IR}}
\newcommand{\cB}{\mathcal{B}}

\newcommand{\beqs}{\begin{equation*}}
\def\beq{\begin{equation}}

\newcommand{\eeqs}{\end{equation*}}
\def\eeq{\end{equation}}

\newcommand{\beqas}{\begin{eqnarray*}}
\newcommand{\beqa}{\begin{eqnarray}}

\newcommand{\eeqas}{\end{eqnarray*}}
\newcommand{\eeqa}{\end{eqnarray}}

\newcommand{\eq}[2]{\begin{equation} #1 \label{#2} \end{equation}}

\newcommand{\al}{\alpha}
\newcommand{\be}{\beta}
\newcommand{\ga}{\gamma}
\newcommand{\de}{\delta}
\newcommand{\om}{\omega}

\newcommand{\blist}{\begin{itemize}}

\newcommand{\elist}{\end{itemize}}

\providecommand{\href}[2]{#2}

\DeclareFontFamily{OT1}{rsfs}{}
\DeclareFontShape{OT1}{rsfs}{m}{n}{ <-7> rsfs5 <7-10> rsfs7 <10->rsfs10}{} 
\DeclareMathAlphabet{\mycal}{OT1}{rsfs}{m}{n}

\usepackage{slashed}

\def\cL{{\cal L}}

\def\cN{{\cal N}}
\def\cM{{\cal M}}

\def\cO{{\cal O}}

\def\cG{{\cal G}}

\def\extd{{\rm d}}

\def\Tr{{\rm Tr}}
\def\TrL2{{\rm Tr}_{L^2}}

\def\atbdry{\Big|_{\partial \cM}}
\def\atbdry0{\Big|_{\partial \cM_0}}
\def\atbdry1{\Big|_{\partial \cM_1}}

\title{AdS/CFT with Flavour in Electric and Magnetic Kalb-Ramond Fields}
\author{Johanna Erdmenger \& Ren\'e Meyer\\ 
Max Planck-Institut f\"ur Physik (Werner Heisenberg-Institut) \\
F\"ohringer Ring 6, 80805 M\"unchen, Germany\\
E-mail: jke@mppmu.mpg.de, meyer@mppmu.mpg.de}
\author{Jonathan P. Shock\\ 
Institute of Theoretical Physics,
 Chinese Academy of Sciences \\  P.O. Box 2735,
 Beijing 100080, People's Republic of China\\
 Email: shock@fpaxp1.usc.es}
\abstract{
We investigate gauge/gravity duals with flavour for which pure-gauge
Kalb-Ramond B fields are turned
on in the background, into which a D7 brane probe is embedded.
First we consider the case of a magnetic field in two of
the spatial boundary directions. We show that at finite temperature, i.e.~in
the AdS-Schwarzschild background, the B field has a stabilizing effect on the
mesons and chiral symmetry breaking occurs for a
sufficiently large value of the B field. Then we turn to  the electric case of
a B field in the temporal direction and one spatial boundary direction. In this case,
there is a singular region in which it is necessary to turn on a gauge field
on the brane in order to ensure reality of the brane action. We find that the
brane embeddings are attracted towards this region.  
Far away from this region, in the weak field case at zero temperature,
we investigate the meson spectrum and find a mass shift
similar to the Stark effect.}
\keywords{{AdS-CFT correspondence, Brane Dynamics in Gauge Theories}}
\preprint{MPP-2007-122}

\begin{document}

\section{Introduction and Summary}\label{sec:intro}
In the last decade we have seen major advances in the understanding
of strongly coupled gauge theories through the AdS/CFT
correspondence and its generalisations. The breaking of
supersymmetry in deformed supergravity backgrounds
and the addition of fundamental matter using brane probes has
allowed a study of QCD-like theories and many measurable quantities
thereof. An example of this is the embedding of probe D7 branes
\cite{hep-th/0205236}. The comparison to real-world QCD is often at the
level of a few percent. In particular, gravity dual constructions describing
spontaneous chiral symmetry breaking have been found in 
\cite{hep-th/0306018,MyersMateos,Sugimoto,Shock}, and are relevant
for the light-quark sector of QCD. The exploration of the phase structure of
this sector is ongoing \cite{hep-th/0306018,Kirsch,
hep-th/0502088,hep-th/0605046,Johnson06,Karch06,Landsteiner,MyersI,MyersII,SinI,SinII,Ghoroku,KarchSeptember,Parnachev}.

In this paper we consider the effect of an external
Kalb-Ramond  B field for D7 branes embedded in
a (dual) gravity background. Our motivation to look at such configurations
originates from the investigations in \cite{AELS,Christoph},
where D7 brane probes were embedded into the Polchinski-Strassler background
\cite{PS}. In this background, a B field of quite complicated structure,
dependent on the radial coordinate of the dual space, is turned on in all six
directions perpendicular to the boundary. In the original work of Polchinski and Strassler \cite{PS} this
has been shown to correspond to mass terms for the adjoint chiral multiplets
in the dual gauge theory. For the theory with added flavour, this implies a
repulsion of the D7 branes by the shell forming in the background due to
the Myers effect \cite{Myerseffect}. This leads to a shift of the meson spectrum induced by the
adjoint masses (for details to second order in the adjoint masses see
\cite{AELS}).

In contrast, in \cite{hep-th/0701001} a pure-gauge magnetic
Kalb-Ramond field was turned on in two spatial directions of the
D3-brane world-volume, parallel to the boundary. This was found to induce chiral symmetry breaking and a
Zeeman splitting of the meson states. This setup was investigated further
in \cite{Filev}. A related approach using both external magnetic and electric fields 
was used in \cite{KarchOBannon,OBannon,Matsuo} to
calculate conductivities and to study the holographic Hall effect
(for related work in 2+1 dimensions see also \cite{Hartnoll}).

In the present paper, we consider both external magnetic and 
electric fields separately. In the electric case, the B field extends into the
temporal direction and one spatial direction parallel to the boundary. We
begin by considering
 the magnetic B field of \cite{hep-th/0701001}
in the AdS-Schwarzschild background dual to a finite-temperature
field theory, and study the phase structure of one fundamental hypermultiplet in
this background by embedding a probe D7 brane in it. We see two competing mechanisms at work: As first
discussed in \cite{hep-th/0306018}, the black hole attracts the D7
brane which bends towards it. For very large values of the quark
mass in units of the temperature, as determined by the UV boundary
value of the D7 embedding, this leads to a very small value of the
quark condensate. Moreover, for brane probes ending on the black
hole, {there is a phase in which mesons are unstable and melt in the $\cN=4$ plasma.} The
melting transition for the mesons occurs when the
D7 brane probe reaches the black hole {horizon}. On the
other hand, as discussed in \cite{hep-th/0701001}, the magnetic B
field leads to spontaneous chiral symmetry breaking, since the quark
condensate is large even at zero quark mass. This is essentially due
to the fact that the magnetic B field has the effect of
repelling the D7 probe from the origin. Here we find that for
sufficiently large B field, the second mechanism is stronger than
the first one and we find spontaneous chiral symmetry breaking even
in the black hole background. A critical line in the $T-B$ phase
diagram illustrates the interplay between the two effects: The
magnetic field acts by repelling the embeddings from the horizon.
Above a critical value for the field strength (at a fixed
temperature) this repulsion is so strong that no embedding, not even
the embedding corresponding to massless quarks, can flow into the
black hole any more - all embeddings are Minkowski 
embeddings for which the meson spectrum is discrete. For fixed mass,
varying the temperature, this is equivalent to stating that for increasing 
magnetic field the critical temperature for the
melting phase transition increases.
We also investigate the meson spectrum for this scenario and find that a
Goldstone mode occurs above the critical B field value, in agreement with
spontaneous chiral symmetry breaking.

Moreover, we also consider the case where
the B field is turned on
in spatio-temporal directions, corresponding to
an external electric field in the gauge theory. 
In this case there
is a singular region with topology $S^5$  where the Dirac-Born-Infeld action
for the D7 brane has a zero and becomes tachyonic.  It is necessary to
 turn on a gauge field on the brane in order to ensure a regular action
beyond the shell where the brane action vanishes,
similar to \cite{KarchOBannon}. The singular shell of vanishing brane action
has an attracting effect on the D7 brane probes, as opposed to the repulsion
observed in the magnetic case. This may be interpreted as {the holographic incarnation of an ionization effect.}

Far away from this region,
i.e.~in the weak field limit at zero temperature, we investigate the meson spectrum both
analytically and numerically and find a mass shift
for the pseudoscalar mesons $\delta M \sim B^2$.
This is very similar to the second-order
Stark effect for atoms in electric fields, where the energy levels of
$s$-orbitals are shifted by an amount proportional to the square of the
applied field strength. For the analytical calculation, we perform  a
perturbative analysis in the external field to lowest order.

{We also consider the case of a general, not necessarily small, electric field strength. After obtaining a regular 
action as in \cite{KarchOBannon} by introducing gauge fields on the brane, we find that both at zero and 
 finite temperature regular embeddings exist which pass
through the shell of vanishing action. Our ansatz for the world volume gauge field corresponds to baryon number densities and  currents in the dual gauge theory, and we are considering the canonical ensemble, in which the baryon number density is fixed. Minkowski embeddings can only exist if the baryon density vanishes \cite{MyersI}, and this is the situation we will be mostly interested in. Besides Minkowski embeddings, we find two different classes at zero temperature: 
Embeddings that reach all the way to the (extremal) horizon of AdS space, and, at quark masses between the Minkowski embeddings and the ones ending at the AdS horizon, embeddings which end in a conical singularity, the latter of which were first noticed in \cite{Johnson2}.
At finite temperature and for vanishing baryon densities,  
we find, in order of decreasing asymptotic quark mass, Minkowski embeddings, conically singular embeddings and those that fall into the black hole. Both at zero and finite temperature, we
expect a first order phase transition between the Minkowski and singular
shell embeddings, which we associate with mesons dissociation due to the external electric field. The physical meaning of the conically singular embeddings, in particular their stability and role in this phase transition, remains to be investigated.
At finite baryon number density and temperature, we find that black hole embeddings exist, covering the whole range of asymptotic quark mass. As Minkowski embeddings are inconsistent in that case, the dissociation phase transition should then occur between different black hole embeddings, similar to the situation in the canonical ensemble without electric field \cite{MyersI}.}

This paper is organized as follows: We begin by considering the general
  ansatz for both magnetic and electric fields in section 2, and comment on
  some aspects of the magnetic case {at zero temperature} in particular. In section 3 we consider
  the magnetic case at finite temperature and show that spontaneous chiral
  symmetry breaking occurs even at finite temperature above a critical value
  of the magnetic field strength. In section 4 we
consider the electric case. In particular, we derive the meson mass shift
which corresponds to the Stark effect. Moreover
we discuss the behaviour of the embeddings for general, not necessarily small, values of the Kalb-Ramond field.
{We conclude briefly in section 5 with a
discussion of the instability which potentially occurs for the electric B-field, 
and differences between the canonical and grand canonical ensemble in the electric case.}
Some calculational details are relegated to two appendices.

While preparing this paper, we became aware of the fact that two papers on {the same subject}
are being prepared
by T.~Albash, V.~Filev, C. Johnson and A.~Kundu \cite{Johnson1,Johnson2}.

\section{Fundamental matter in external fields  at zero temperature}\label{sec:fundmatzeroT}

In \cite{hep-th/0701001} a pure gauge magnetic Kalb-Ramond field was
added to the $AdS_5\times S^5$ geometry and quenched fundamental
matter was included using a D7 brane probe. The D7 brane embeddings
were calculated and it was found that chiral symmetry is
spontaneously broken, as indicated by a non-zero quark condensate at
vanishing quark mass. The spectrum of mesons was also calculated
both for small and large  magnetic field strength and the Goldstone
mode was found to satisfy the Gell-Mann-Oakes-Renner relation
\cite{GellMann:1968rz}.

In this section we review the results of \cite{hep-th/0701001}.
Moreover we simultaneously consider
a new ansatz for the Kalb-Ramond field, corresponding to an electric field.

\subsection{General ansatz}\label{sec:generalansatz}

We deal in parallel with the magnetic and electric ans\"atze for
the Kalb-Ramond field.

The background of interest is the pure AdS geometry, given by
\begin{equation}\label{eq:pureads}
ds^2=\frac{\omega^2}{R^2}\left(dx_0^2+d\vec{x}^2\right)+\frac{R^2}{\omega^2}\left(d\rho^2+\rho^2d\Omega_3^2+dL^2+L^2d\Phi^2\right).
\end{equation}
Here the AdS radial coordinate is given by $\omega^2 = \rho^2+L^2$.
For the $S^3$ we use Hopf coordinates
\begin{equation}  \label{Hopf}
d\Omega_{3}^2 = d\psi^2+\cos^2\psi d\beta^2+\sin^2\psi d\gamma^2 \, .
\end{equation}
In addition, the background involves the usual four-form and constant dilaton
\begin{gather}
g_{s}C_{(4)} = \frac{\om^4}{R^4}dx^0\wedge dx^1\wedge dx^2 \wedge
dx^3\, , \quad e^{\phi_\infty}= g_s \, , \quad R^4 = 4\pi g_{s}N \alpha'^2 \,
.\label{eq:C4dilaton}
\end{gather}
There are two obvious choices of a pure gauge ansatz for the
Kalb-Ramond fields\footnote{A third choice, as pointed out to us by
P. Aschieri, would be a light-like B field. It has the advantage
that for spacelike (i.e. magnetic) or lightlike $B$, there is a
decoupling limit leading to a noncommutative field theory
\cite{hep-th/0106281,hep-th/9908142}, while in the electric
or timelike case the best one can do is to decouple the closed
string modes \cite{hep-th/0005040,hep-th/0005048,hep-th/0005073}. This then
corresponds to non-commutative open string theory.},
\begin{equation}
B_{(2)m}=B dx_2\wedge dx_3, \hspace{1cm} B_{(2)e}=B dx_0\wedge dx_1.
\label{eq:KRAnsatz}
\end{equation}
We label these the magnetic and electric ansatz, respectively, as,
after embedding a D7 brane in the background and trading the constant $B$ field for the $U(1)_F$ gauge field on
the brane via $F_{\mu\nu} = - \frac{B_{\mu\nu}}{2\pi\alpha'}$, the
two choices correspond to constant electric and magnetic field
strengths. Both these pure-gauge ans\"atze are solutions to the IIB
supergravity equation of motion (see eg.~\cite{PS}). Provided that $C_2=0$, which ensures the absence of a boundary term $\extd(C_2\wedge B)$, the B field does not deform the $AdS_5\times S^5$.

{In this work, we will 	always choose static gauge $\xi^a = (x^\mu, \rho, \psi, \be, \ga)$,} for the D7 brane,
and parametrize its embedding by an ansatz $L=L(\rho)$, $\Phi=0$, thus preserving the rotational symmetry of the $S^3$ wrapped by the D7 brane. The rotational symmetry perpendicular to the D7 brane will be broken for nonzero embeddings $L$. 
In the background \eqref{eq:pureads} and for the two B fields \eqref{eq:KRAnsatz}, the DBI action is given by 
\begin{equation}\label{eq:Lmande}
{\cal
L}_{m,e}=-\frac{T_7}{g_s}\rho^3\sin\psi\cos\psi\sqrt{1+L'^2}\sqrt{1\pm\frac{R^4B^2}{\left(\rho^2+L^2\right)^2}}
\, ,
\end{equation}
where the positive (negative) sign corresponds to the magnetic
(electric) ansatz, and $\psi$ is one of the Hopf coordinate angles
defined in (\ref{Hopf}). The Wess-Zumino part of the D7 action does not contribute, as $P[C_4]\wedge B_{(2)m,e} = 0$ and the pullback of the magnetic dual of $C_4$ vanishes, $P[\tilde{C}_4]=0$. 
The D7 brane embedding is found by solving
the Euler-Lagrange equation for $L(\rho)$, 
\eq{0=\partial_\rho\left( \frac{\rho^3
L' \sqrt{1 \pm \frac{B^2 R^4}{(\rho^2 +
L^2)^2}}}{\sqrt{1+L'^2}}\right) \pm \frac{2B^2 R^4 \rho^3 L
\sqrt{1+L'^2}}{(\rho^2 + L^2)^3 \sqrt{1 \pm \frac{B^2 R^4}{(\rho^2 +
L^2)^2}}}\, ,}{eq:embedBelmag} 
 which is a scalar from the point of view of the world volume field theory.
 In both the magnetic and electric
cases, the UV (i.e. large $\rho$) behaviour of the embeddings is
given by
\begin{equation}\label{eq:UVas}
{L}({\rho}) \sim {m}+\frac{{c}}{{\rho}^2}\,,
\end{equation}
i.e.~the embeddings asymptote to the
pure $AdS_5\times
S^5$ solution $L=m$ for $\rho \rightarrow \infty$. The quantity $m$ is proportional to the quark mass, $m_q = m/2\pi\alpha'$, while $c$ is related to the chiral condensate via\footnote{Because of supersymmetry, $c$ would also contain squark terms. We assume in this work that the squarks have zero vacuum expectation value \cite{MyersI}.} $\langle \bar{q} q \rangle = c/(2\pi\al')^3$.
As discussed in \cite{hep-th/0701001}, supersymmetry is broken on the brane,
though  not in the background.

\subsection{Magnetic Kalb-Ramond field at zero temperature}\label{sec:magneticzeroT}

We now review the results of \cite{hep-th/0701001} for the magnetic
case at zero temperature. Some examples for the
embedding in the magnetic case with varying IR boundary \
are shown in figure \ref{fig:B23flows}. For completeness we have
also shown solutions with negative quark mass as fixed by the UV
asymptotic behaviour. For small quark mass, the embeddings
intersect the $\rho$ axis, in some cases several times. We will show
that these solutions describe a well-behaved renormalization group
flow, in contrast to the argument in \cite{hep-th/0701001}.
Nevertheless these solutions are ruled out by an energy argument
which shows that they are not the lowest-energy configurations.
\FIGURE[ht]{
\includegraphics[width=16cm,clip=true,keepaspectratio=true]{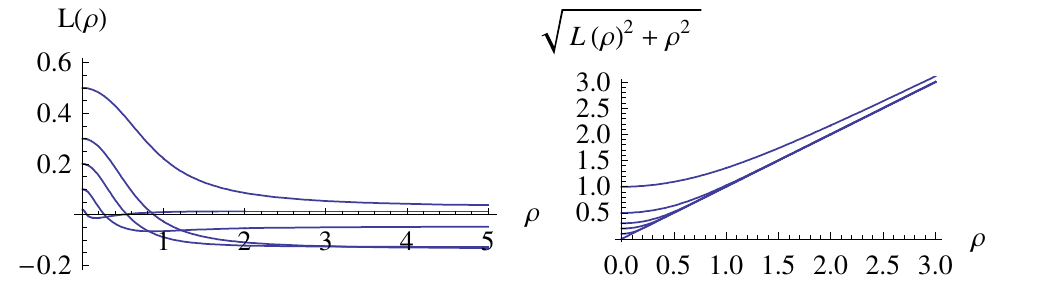}
\caption{D7 brane embeddings in $AdS_5\times S^5$ with external magnetic field.}
\label{fig:B23flows}}
Although the D7 brane does cross the $\rho$ axis multiple times for
small quark masses, it does not indicate a
multiple intersection with the D3-brane stack. The distance of the
D7 brane from the origin of the $(\rho,L)$ plane is monotonically
decreasing as the solution flows towards the IR. This can be seen in the right
hand graph of figure \ref{fig:B23flows}.

In the normalization of \cite{hep-th/0701001}, for which we perform
a rescaling $\rho \rightarrow R\sqrt{B}\rho$, $L\rightarrow R\sqrt{B}L$, the
equation of motion is given by \eq{0=\partial_\rho\left(
\frac{\rho^3 L' \sqrt{1 + \frac{1}{(\rho^2 +
L^2)^2}}}{\sqrt{1+L'^2}}\right) + \frac{2 \rho^3 L
\sqrt{1+L'^2}}{(\rho^2 + L^2)^3 \sqrt{1 + \frac{1}{(\rho^2 +
L^2)^2}}}\, .}{eq:embedBij} 
The UV asymptotics of the solutions is
\begin{gather}
L(\rho)=\tilde{m}+\frac{\tilde{c}}{\rho^2} \, ,
\quad \tilde{m}=\frac{2\pi\alpha'm_q}{R\sqrt{B}} \, , \quad
\tilde{c}=\left<\bar{q}q\right>\frac{\left(2\pi\alpha'\right)^3}{R^3B^{\frac{3}{2}}}
\, . \label{eq:UVT0}
\end{gather}
The extra factor of $B$ in the normalization is convenient in the
zero temperature case. Using (\ref{eq:UVT0}),  we can extract the
rescaled mass and condensate from the numerically determined
D7 embeddings.  It was found in \cite{hep-th/0701001} that
$\tilde{c}(\tilde{m})$ shows a self-similar spiral behaviour.  This indicates
that for a given quark mass, there may be more than one solution. It
is thus important to understand which of the solutions is physical.

It was stated in \cite{hep-th/0701001} that the inner arms of the
spiral intersect with the D3 branes multiple times, and thus only the outer
portion of the spiral in the lower right quadrant of {figure
\ref{fig:spiral}} is physical. We now show that this conclusion may
be found through simple energy considerations.
\FIGURE[ht]{
\includegraphics[width=14cm,clip=true,keepaspectratio=true]{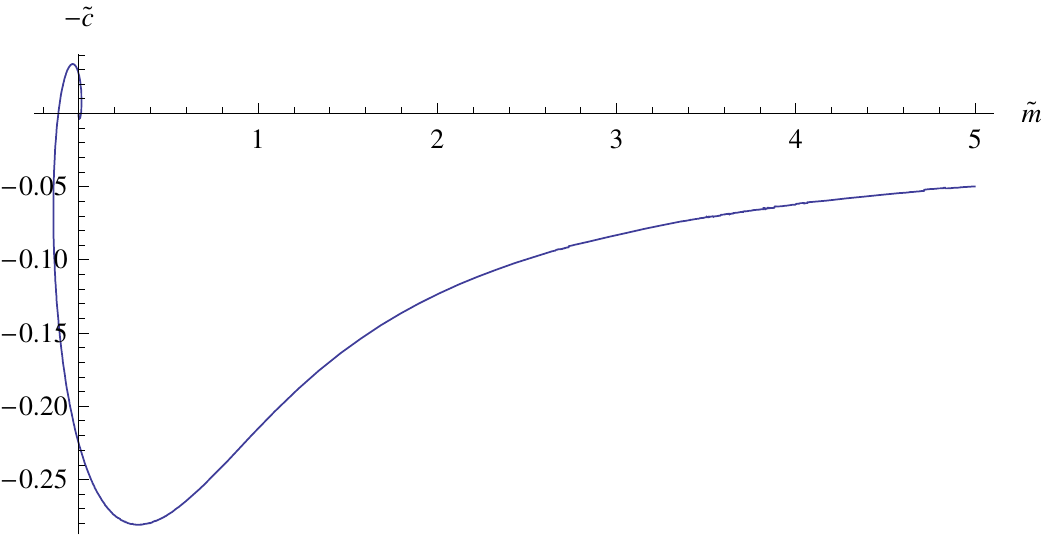}
\caption{$-\tilde{c}(\tilde{m})$ for the magnetic ansatz at $T=0$.
}
\label{fig:spiral}}
In order to determine which of the degenerate solutions is the
physical one, we calculate the energy of each solution. The energy of each one 
has a UV divergence which must be removed by an appropriate
normalisation. For the purposes of our calculation, it is sufficient
simply to impose a cutoff $\Lambda$, which is sent to infinity at the end, and subtract a reference
solution \footnote{As the energy for our static D7 brane
  configuration is just the negative of the action, we drop here all volume factors arising
  from integration over $(x^\mu, \psi, \beta, \gamma)$. Note that here we are not considering
a holographic renormalisation and regularisation which is necessary for
the correct calculation of the free energy and thermodynamic quantities \cite{Johnson1}. In the coordinate system used in this work, no  counterterms including $m$ or $c$ are necessary to cancel the large volume divergence $\propto \Lambda^4$, which is achieved by our subtraction method as well.
},
\begin{equation}
E_{norm}=\int_{\rho_0}^\Lambda
d\rho\rho^3\sqrt{1+L_0'(\rho)^2}\sqrt{1\pm\frac{B^2R^4}{\left(\rho^2+L_0(\rho)^2\right)^2}}-E_{ref}\
,\label{eq:Enorm}
\end{equation}
where
\begin{equation}
E_{ref}=\int_{\rho_0}^\Lambda
d\rho\rho^3\sqrt{1+L_{ref}'(\rho)^2}\sqrt{1\pm\frac{B^2R^4}{\left(\rho^2+L_{ref}(\rho)^2\right)^2}}\
.\label{eq:Eref}
\end{equation}
Here $L_0$ is the classical solution to the embedding equation of motion.
The lower integration limit $\rho_0$ is zero for the magnetic case at zero temperature, as in that case the D7 brane fills the whole range $ 0\le \rho < \infty$.
We use the physical $m=0$ solution as a reference solution. 
 This is just a convenient choice, as a shift of the normalized energy does not change the relative energies between physical and unphysical embeddings.

The $-\ctilde(\mtilde)$ spiral in figure \ref{fig:spiral} cuts the
$\tilde{m}=0$ axis an infinite number of times, dividing the spiral
in different branches in each quadrant of the
$(\mtilde,\ctilde)$-plane. The physical embeddings are those which
have the lowest energy according to \eqref{eq:Enorm}. Figure
\ref{fig:Ecm} shows the $-\ctilde(\mtilde)$-spiral (green line)
along with the corresponding normalized energy $E_{norm}(\mtilde)$
(blue line). The dashed lines link $\mtilde=0$ points on the
condensate curve which correspond to massless embeddings, with the
corresponding points on the normalized energy curve. The numbers
next to the dashed lines are the values of the normalized energies
for these massless embeddings. We find that the lowest energy
configuration for zero quark mass is 
the one where the lowest branch of the condensate curve in the
bottom right quadrant of figure \ref{fig:Ecm} intersects the
$-\ctilde$-axis. It is also clear from figure \ref{fig:Ecm} that the
energy for the embeddings on this branch is indeed smaller than for
all the other branches. The lowest lying branch in the bottom right
quadrant of figure \ref{fig:Ecm} thus corresponds to  the physical
embeddings, while the other branches have to be considered
unphysical. As this physical branch admits a  non-zero quark
condensate $-\ctilde(0)$ at zero quark mass, spontaneous breaking of
the $U(1)$ chiral symmetry is possible in the presence of the
magnetic Kalb-Ramond field.

\FIGURE[ht]{
\includegraphics[width=14cm,clip=true,keepaspectratio=true]{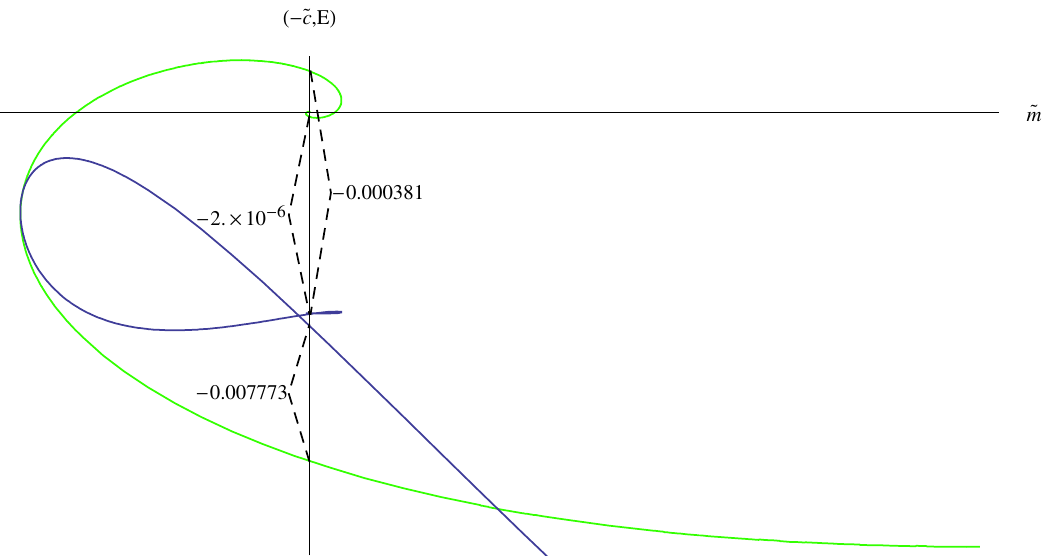}
\caption{$-\tilde{c}(\tilde{m})$ for the magnetic ansatz (green line)
and $E_{norm}(\tilde{m})$ (blue line). The numbers correspond to
energies, and the dashed lines link corresponding points on the two
curves. The origin of two plots are separated vertically by around
0.13 units. Note that the origin of the $\ctilde$ axis does not coincide with the origin of the $E_{norm}$ axis. See text for detailed explanation.
} \label{fig:Ecm}}


\section{{Magnetic} Kalb-Ramond field at finite temperature}\label{sec:magneticfiniteT}

The embedding of D7 brane probes into the AdS-Schwarzschild black
hole background, dual to a finite-temperature field theory, has
first been studied in \cite{hep-th/0306018}. In the black hole
background there is no spontaneous chiral symmetry breaking by a
quark condensate, since the quark condensate vanishes for zero quark
mass. On the other hand, there is an interesting first order phase
transition \cite{Kirsch} when the embedded D7 brane reaches the
horizon. This has been studied in further detail by many authors
\cite{hep-th/0502088,hep-th/0605046,Johnson06,Karch06}, including the effect of a baryon chemical potential and of a finite baryon number density \cite{SinI,SinII,Ghoroku,KarchSeptember}. The embeddings which terminate before
reaching the horizon do so because the $S^3$ which they wrap shrinks to
zero size, as discussed in \cite{hep-th/0205236}. The phase
transition
 corresponds to the transition from the mesonic to the
melted phase \cite{Landsteiner}. In the former there is a
discrete meson spectrum with a mass gap, whereas in the latter the
spectrum becomes continuous. The mesons melt at this phase
transition due to the interaction with the hot $\cN=4$ plasma. {Subsequently we will refer to D7
embeddings reaching the black hole horizon as black hole embeddings or as ``in a melted phase'', while
those which do not are named mesonic or Minkowski ones.}

We now consider the effects of the Kalb-Ramond field
\eqref{eq:KRAnsatz} in the AdS-Schwarz\-schild background
 that is dual to {a} finite temperature field theory.
The metric in Minkowski signature is given by
($\omega^2=\rho^2+L^2$)
\begin{equation}\label{eq:AdSS}
ds^2=\frac{\omega ^2}{2R^2} \left(\frac{\extd \vec{x}^2
   \left(b^4+\omega ^4\right)}{\omega ^4}-\frac{\text{dt}^2 \left(\omega ^4-b^4\right)^2}{\om^4 \left(b^4+\omega ^4\right)}\right)+\frac{R^2}{\omega ^2}\left(\text{d}L^2+\text{d$\rho $}^2+\text{d$\Phi $}^2
L^2+\rho ^2 \text{d$\Omega $}_3^2\right)\,.
\end{equation}
The dilaton is constant for the black (i.e. non-extremal) D3 brane
solution, whose near-horizon geometry is just \eqref{eq:AdSS}
\cite{hep-th/9803131}. {The temperature of the dual field theory
is given by the Hawking temperature of the black hole}
\begin{equation}
T=\frac{b}{\pi R^2} \,.
\end{equation}

With the same static gauge as before, the DBI
Lagrangian for a D7 brane probe in this background reads for the magnetic (upper sign) and electric (lower
sign) ansatz
\begin{align}
{\cal L} = & -\frac{T_7}{g_s}\cos\psi\sin\psi \frac{\tilde{\rho} ^3 \left((\tilde{L}^2+\tilde{\rho}^2)^2\pm
1\right)}
{4 (\tilde{L}^2+\tilde{\rho}
   ^2)^4} \times 
   \nonumber\\ & \qquad \times
\sqrt{\left(\left(1\mp(\tilde{L}^2+\tilde{\rho}
^2)^2\right)^2\mp\tilde{B}^2
   (\tilde{L}^2+\tilde{\rho} ^2)^2\right) (\tilde{L}'^2
+1)} \,. \label{eq:BTans}
\end{align}
Again, we have performed  rescalings to dimensionless quantities
$(L,\rho,B)=(b\tilde{L},b\tilde{\rho},\frac{\tilde{B}b^2}{2R^2})$,
such that the horizon is at $\tilde{\rho}^2+\tilde{L}^2=1$.


Let us now turn to the magnetic case. We numerically solve the Euler-Lagrange equation for $\tilde{L}(\tilde{\rho})$ 
obtained from the Lagrangian (\ref{eq:BTans}). 
It is convenient to introduce a rescaled B field 
\begin{equation}
\hat{B}=\frac{
B\lambda}{2\pi^2R^2m_q^2} \, ,
\end{equation}
as well as an appropriate dimensionless quark mass and condensate
\begin{gather}
m_q = \frac{1}{2}\sqrt{\lambda} T \tilde{m}\, , \qquad \tilde{T} = \frac{1}{\tilde{m}} \,,\qquad 
\left<\bar{q}{q}\right> = -\frac{1}{8}\sqrt{\lambda}N_cT^3 \tilde{c}
\,.
\end{gather}
Here $\lambda = g_s N_c$ is the 't Hooft coupling.

The numerical results are plotted in figure \ref{fig:BandTflowsv}
for increasing values of the magnetic field at
a fixed temperature (or equivalently fixed Schwarzschild radius).
\FIGURE[ht]{
\includegraphics[width=17cm,clip=true,keepaspectratio=true]{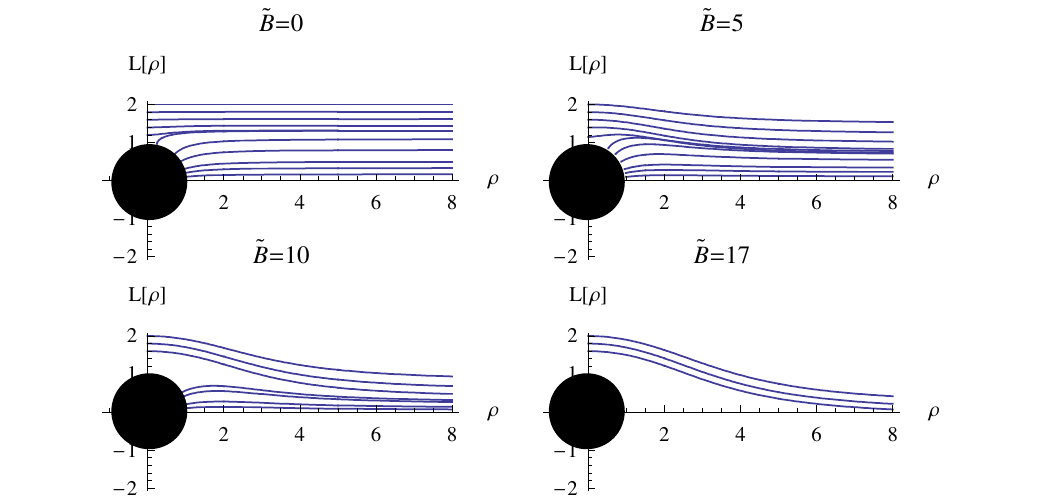}
\caption{Increasing values of $\tilde{B}$ for
fixed $T$ show the repulsive nature of the magnetic Kalb-Ramond field.
We see that for
large enough $\tilde{B}$, the melted 
phase is never reached, and the chiral symmetry is spontaneously broken.}
\label{fig:BandTflowsv}}
We see
that the increasing
external magnetic field repels the branes from the horizon
more and more, until there are no black hole solutions any more.
This is exactly the point where the melted phase disappears, at a critical value
$\tilde{B}_{crit} \approx 16$. In fact, above this critical value, there
is spontaneous
chiral symmetry breaking, since the lowest-energy solution at quark
mass $\tilde
m=0$ is a mesonic one and has a condensate $\tilde c > 0$.
On the other hand, in the case where the zero quark mass solution
reaches the horizon and therefore corresponds to melted mesons, this
solution is given by $L(\rho)=0$ and thus no condensate develops and
no spontaneous chiral symmetry breaking occurs.


\FIGURE[ht]{
\includegraphics[width=15cm,clip=true,keepaspectratio=true]{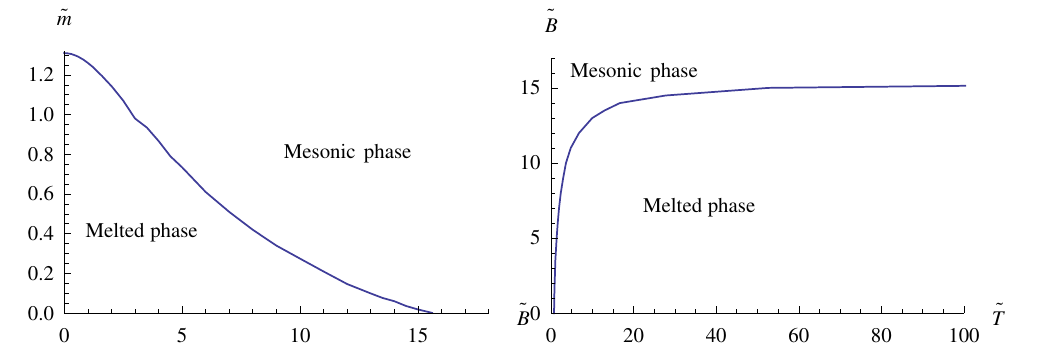}
\caption{Phase diagram in the $(\tilde{m},\tilde{B})$ and
$(\tilde{T},\tilde{B})$ planes for the
magnetic Kalb-Ramond field. The graph on the left shows the largest
quark mass for a given $\tilde B$ for which the embedding still reaches
the black hole. In the graph on the right,
the critical line reaching a constant
$\tilde{B}$ at large temperature corresponds to a quadratic
dependence $\hat{B}_{crit}\sim T^2$ at large temperature.}
\label{fig:condeconfpbs}}
The phase diagram is depicted in figure \ref{fig:condeconfpbs}.
{On the left hand side we plot the largest quark mass $\tilde m$ for given
$\tilde B$ for which the embedding still reaches the
horizon. The  temperature is fixed.} We see that above the critical value
$\Btilde_{crit}  \approx 16$, there are no more black hole embeddings
reaching the black hole horizon.
On the right hand side of figure \ref{fig:condeconfpbs},
we consider the same phase diagram for fixed $\tilde
m$ while varying $\tilde T$.
We see that for large $\tilde{T}$, the critical value
$\tilde{B}_{crit}$ tends to a constant value around 16. Because
$\tilde{B}$ is explicitly a function of $T$, the appropriate
dimensionless quantities to plot on the $(T,B)$ phase diagram are
$(\tilde{T},\hat{B})$. Because $\tilde{B}_{crit}$ becomes
constant for large $\tilde{T}$, we see that $\hat{B}_{crit}$ behaves as
$\tilde{T}^2$ for large $\tilde{T}$, corresponding to large
temperature at fixed quark mass.

The condensate as a function of the
mass for different $\tilde B$ is shown in
figure \ref{fig:cvsmxx}, which can be found at the end of this work: The blue
part of the curve corresponds to the mesonic phase, while the green
part at small quark masses corresponds to the melted phase.
Increasing the magnetic field strength lowers the critical quark
mass at which the first order phase transition occurs, until zero
quark mass is reached at the critical value $\tilde{B}_{crit}$. At
this point there is no melted phase any more, but a non-zero
condensate at zero quark mass indicates spontaneous chiral symmetry
breaking in the dual gauge theory. It should be noted that for a
given quark mass there may be several possible embeddings. Just as in the case
of pure finite temperature or pure magnetic field, the physical
solutions can be found by energy considerations.


Yet another graph showing clearly the onset of spontaneous chiral symmetry
breaking at the critical value of the magnetic B field is figure \ref{fig:cB2},
in which the condensate at zero quark mass is plotted versus the external
magnetic field. We observe that the phase transition is first order{, as there is a jump in the order parameter $\ctilde$.}
\FIGURE[ht]{
\includegraphics[width=10cm,clip=true,keepaspectratio=true]{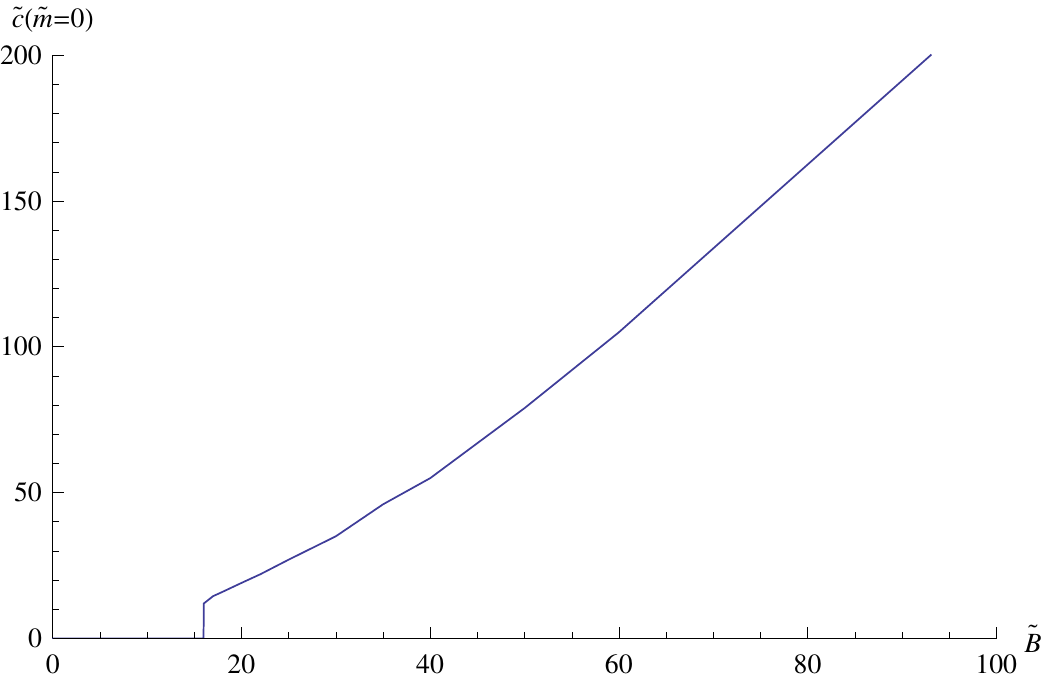}
\caption{Condensate for the lowest-energy embedding at zero quark mass $\ctilde(0)$ as function of the
magnetic field $\tilde B$. There is a first-order phase
transition at which spontaneous chiral symmetry breaking occurs.
}
\label{fig:cB2}}

{We also calculate meson masses for the Minkowski embeddings in the magnetic finite
temperature background by solving the eigenvalue problem for the $\Phi$-fluctuations. These
correspond to pseudoscalar mesons, and we restrict ourselves to states with zero $S^5$
angular momentum $l=0$. This calculation was performed at zero temperature in
\cite{hep-th/0701001}, where it was found {(see also appendix~\ref{app:B})} that it is possible to decouple the gauge field
fluctuations on the brane from the fluctuations of the embedding coordinate $\Phi$, if the
latter depends only on the coordinates $(x^2,x^3)$. These modes
 can be thought of as representing a  meson in the euclidean $(x^2,x^3)$-plane
\cite{hep-th/0701001}. We follow the same strategy and thus use
the ansatz}
\begin{equation} \label{eq:magfluc}
\phi(\rho,x^2,x^3)=h(\rho)e^{-ik_2x_2-ik_3x_3} \,.
\end{equation}
The Lagrangian for the fluctuations reads
\begin{gather}\nonumber
{\cal L} = -\frac12 \frac{T_7}{g_s}\cos\psi\sin\psi
\rho^3\sqrt{g_{\rho\rho}^4
g_{x^1x^1}^3g_{tt}\left(\frac{B^2}{g_{x^1x^1}^2}+1\right)
 \left(L'^2+1\right)} \times \\\hspace{1cm} \times
 g_{\phi \phi} \left(
\frac{
  g_{x^1x^1} \left( (\partial_2\phi(\rho,x^2,x^3))^2 + (\partial_3\phi(\rho,x^2,x^3))^2\right)}{\left({B^2}+{g_{x^1x^1}^2}\right)
}+\frac{ (\partial_\rho\phi (\rho,x^2,x^3))^2}{ g_{\rho\rho}
   \left(L'^2+1\right)}\right) \, ,\label{eq:fluctmagfiniteT}
\end{gather}
where the metric is given by \eqref{eq:AdSS}.

\FIGURE[t]{
\includegraphics[width=10cm,keepaspectratio=true]{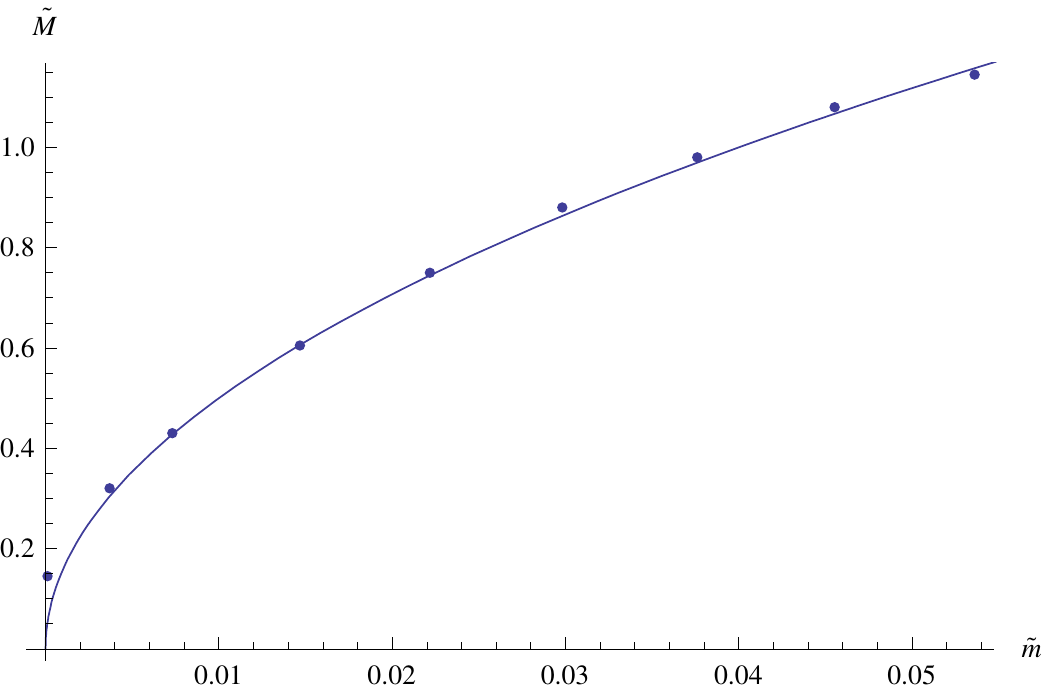}
\caption{{Mass of the Goldstone boson for $\Btilde=16$. The solid line is the fitted curve $\tilde{M}=5\sqrt{\tilde{m}}\,,$ showing the typical  Gell-Mann-Oakes-Renner behaviour.}}
\label{fig:GMOR}
}

The meson masses are obtained from
\begin{equation}
M^2 = - k_2^2 - k_3^2 \, .
\end{equation}
The meson spectrum for fixed temperature and different B field values is plotted in figure \ref{fig:spectra}, where the dashed lines are the pure AdS result $\Mtilde = 2 \mtilde\sqrt{(n+1)(n+2)}$, cf.~\cite{hep-th/0304032}.
The mesons in the magnetic case are heavier than their pure AdS counterparts, from which we conclude that the magnetic B field 
lowers the binding energy $E_b<0$ and thus has a confining effect on the mesons. 
 We find that above the critical magnetic field strength
$\Btilde_{crit}\approx 16$, the lowest-lying meson state becomes
massless for zero quark mass and thus is the Goldstone mode of
chiral symmetry breaking. {Figure \ref{fig:GMOR} zooms into the zero mass region for this state, calculated for
$\Btilde=16$. The solid curve, $\tilde{M}=5\sqrt{\tilde{m}}$, is obtained by fitting to the data points, and shows 
that our identification of this state with the Goldstone mode of chiral symmetry breaking is correct, as it satisfies the 
Gell-Mann-Oakes-Renner relation \cite{GellMann:1968rz}. The deviation of the data point for the lowest reduced quark mass $\tilde{m}$ is due to numerical artefacts.
}


\section {Electric Kalb-Ramond field}\label{sec:electric}

\subsection{Electric field at zero temperature}\label{sec:electriczeroT}

We now turn to the case when the Kalb-Ramond B field is switched on in the
spatio-temporal directions, corresponding to an external electric field.
There is a subtle but important difference between the
Lagrangians for  the magnetic and electric ansatz
 \eqref{eq:Lmande}:  In the case of the
electric field, there is a zero of the action at $\omega^2 \equiv
\rho^2+L^2=BR^2$. We name the {five-sphere} defined by this equation
the {\it singular shell}. Generically, we find that 
in contrast to the magnetic case where
the external field has a repulsive effect, now the singular shell is
attracting the D7 solutions, in a sense similar to the black hole metric.

For values of the radius $\omega$ inside the singular shell, the
DBI Lagrangian (\ref{eq:Lmande}) 
becomes imaginary, which indicates a tachyonic instability
\cite{Burgess,Nesterenko,BachasPorrati,Bachas}. We stabilize 
the D7 configuration by switching on a compensating gauge field on the D7
brane near and inside the singular 
shell, along the lines of \cite{KarchOBannon,OBannon}: We switch on components of the gauge field's Faraday tensor which correspond holographically to a finite baryon number density and a baryon number current expectation value $\langle J_x \rangle$. 
Demanding the DBI action to be real throughout ten-dimensional spacetime then imposes a regularity constraint relating number density and current.\footnote{In fact, for the reality of the DBI action it is enough to introduce the magnetic component $F_{\rho x}$. We however also consider the electric component $F_{\rho t}$, which corresponds to introducing a chemical potential and a finite baryon number density on the brane.}

Using the ansatz of \cite{KarchOBannon},
\eq{A_x = -f(\rho)\,,\quad A_t = -g(\rho)\,,}{eq:gaugeansatz}
the action for the D7 brane in the external electric B field takes the
form\footnote{One may check that for the RR four-form potential \eqref{eq:C4dilaton}, the Chern-Simons part of the D7-brane action only contributes through the gauge field strength components $F_{\hat{a}\hat{b}}\,,\,\{\hat{a},\hat{b}\}\in\{\rho,\psi,\beta,\gamma\}\,.$ The coupling to the magnetic dual four-form $\tilde{C}_4$ vanishes because $P[\tilde{C}_4] = 0$ for our choice of the embedding. Furthermore, we are suppressing here the customary factor of $2\pi\alpha'$ in front of the Faraday tensor. It can easily be restored.}
\beqa
S_{D7} &=& 
- \cN \int\extd\rho  \rho^3 
\sqrt{f'(\rho)^2 - g'(\rho)^2 + \left( 1 - \frac{B^2 R^4}{(\rho^2+L^2(\rho))^2}\right)(1+L'^2)}\,.
\label{eq:D7withgaugefield}
\eeqa
The factor of $2\pi^2$ is just the volume of a unit $S^3$, and $\cN = \frac{T_7}{g_s} Vol(\mathbb{R}^4) 2 \pi^2$. As
 the action depends on the $\rho$-derivatives of $A_x$ and
$A_t$ only through the field strength \cite{KarchOBannon}, the Euler-Lagrange equations for $A_t$ and $A_x$ demand the existence of two conserved quantities,
\beqa\label{eq:conservedD}
\delta A_t: D & = & \frac{\cN \rho^3 g'(\rho)}{\sqrt{f'(\rho)^2 - g'(\rho)^2 + \left( 1 - \frac{B^2 R^4}{(\rho^2+L^2(\rho))^2}\right)(1+L'^2)}}\,,\\\label{eq:conservedB}
\delta A_x: \cB & = & -\frac{\cN \rho^3 f'(\rho)}{\sqrt{f'(\rho)^2 - g'(\rho)^2 + \left( 1 - \frac{B^2 R^4}{(\rho^2+L^2(\rho))^2}\right)(1+L'^2)}}\,.
\eeqa
The field theory interpretation of these quantities is given in
 \cite{KarchOBannon}: $D$ corresponds to a finite baryon number density
 $<J_t>$, and $\cB$ to a current in $x$-direction $<J_x>$. Inverting 
 \eqref{eq:conservedD} and \eqref{eq:conservedB} yields 
\beqa
f'(\rho) & = & + \cB \frac{\sqrt{1+L'^2(\rho)}\sqrt{1 - \frac{B^2 R^4}{(\rho^2+L^2(\rho))^2}}}{\sqrt{\cN^2\rho^6 + D^2 - \cB^2}}\,, \label{eq:fprimeBD}\\
g'(\rho) & = & - D \frac{\sqrt{1+L'^2(\rho)}\sqrt{1 - \frac{B^2 R^4}{(\rho^2+L^2(\rho))^2}}}{\sqrt{\cN^2\rho^6 + D^2 - \cB^2}} \label{eq:gprimeBD}\,.
\eeqa
Using these solutions, as in \cite{KarchOBannon}  we now perform
 a Legendre transform eliminating $f$ and $g$ completely from the action and replacing them by the conserved quantities $\cB$ and $D$,
\beqa
\bar{S}_{D7} &=& S_{D7} - \int \extd \rho \left(g'(\rho)\frac{\de S_{D7}}{\de g'(\rho)} + f'(\rho)\frac{\de S_{D7}}{\de f'(\rho)}  \right)\nonumber\\
&=& - \int \extd \rho \sqrt{(\cN^2\rho^6 + D^2 - \cB^2)(1+L'^2)\left(
1 - \frac{B^2 R^4}{(\rho^2+L^2(\rho))^2}
\right)}\,.\label{eq:SD7Legendre} \eeqa We see that it is possible to obtain
an action which is real everywhere by demanding that at the
point $\rhoIR^2 + L(\rhoIR)^2 = BR^2$, where the last bracket in the
square root of \eqref{eq:SD7Legendre} changes sign,
the first bracket therein should also change sign. This implying a
relation between the two conserved quantities and the position where the brane hits the singular shell, 
 \eq{\cB^2 =
\cN^2\rhoIR^6 + D^2\,.}{eq:BDrelation} Reinserting this relation
into \eqref{eq:SD7Legendre}, we find \eq{\bar{S}_{D7}' = - \cN
\int\extd\rho \sqrt{(\rho^6-\rhoIR^6) (1+L'^2)\left( 1 - \frac{B^2
R^4}{(\rho^2+L^2(\rho))^2} \right)}\,.}{eq:SD7Legendre2} 
In the finite temperature
AdS-Schwarzschild background \eqref{eq:AdSS} the corresponding
expressions read ($\LIR = L(\rhoIR)$)
\begin{eqnarray}
\omega_{IR}^2&=&\rhoIR^2 + \LIR^2 = \frac{BR^2}{2}+\frac{1}{2}\sqrt{4b^4+B^2R^4}\, , \\\label{eq:BDrelationT}
\cB^2 & = & \frac{(\omIR^4-b^4)^2}{(\omIR^4+b^4)^2}D^2+\frac{\cN^2 B^2 R^4 (\omIR^4+b^4)\rhoIR^6}{\omIR^{8}}  \, ,\\
\bar{S}_{D7}' & = & -\int d\rho \sqrt{a\left(\cN^2\left(\rho^6-\frac{b_{IR}c_{IR}}{bc}\rhoIR^6\right) + D^2\frac{b-b_{IR}}{bc}\right)}\,,
\nonumber\\
\text{with}&&\nonumber \\
a & = &(1+L'^2)\left(1+\frac{b^4}{\om^4}\right)^2\left( \left(1-\frac{b^4}{\om^4}\right)^2-\frac{B^2R^4}{\om^4}\right)\,, \nonumber\\
b & = & \left(1-\frac{b^4}{\om^4}\right)^2 \left(1+\frac{b^4}{\om^4}\right)\,,\quad c = \left(1+\frac{b^4}{\om^4}\right)^3\,.
\label{eq:SD7Legendre2T}
\end{eqnarray}
At finite temperature the singular shell always resides outside the horizon at $\om^2_{\text{hor}} = b^2$. 
There is an important difference between the cases of finite and zero
temperature: While \eqref{eq:SD7Legendre2T} does, after Legendre
transformation, explicitly depend on $D$, the zero temperature action
\eqref{eq:SD7Legendre2} does not. Both are, however, real, over the whole
range $\rho \in [0,\infty)$ \textit{per constructionem}. The embeddings in the zero temperature case thus will not depend on the baryon number density, but the finite temperature embeddings will: The choice of $D$ at finite temperature influences the asymptotic values of the quark masses for the D7 branes falling into the horizon, i.e. sets an energy scale.

\subsubsection{Embeddings }

We begin by studying  the D7 brane embeddings at zero temperature and with $D=0$, which are depicted in figure~\ref{fig:D7flowszero}. For Minkowski embeddings (depicted in blue), which do not reach the singular shell but flow all the way to $\rhoIR = 0$,  the gauge field can be turned off consistently: $A_x=A_t=0$. There also are embeddings which do intersect the singular shell (depicted in green): Some of these flow towards the origin $L=\rho=0$, i.e. the location of the AdS horizon, while others end in a conical singularity at $\rho=0$. For the singular shell embeddings, the infrared value $\rhoIR$ is given by $\rhoIR^2 + \LIR^2 = BR^2$, where $\LIR$ is the value of $L$ at which they intersect the singular shell.

The action \eqref{eq:SD7Legendre2} is real everywhere, and reduces to \eqref{eq:Lmande} with the lower sign for the Minkowski embeddings, for which $\rho_{IR}=0$. 
The Euler-Lagrange equation for $L(\rho)$ as derived from \eqref{eq:SD7Legendre2} reads
 \eq{\partial_\rho\left( \frac{\sqrt{\rho^6-\rhoIR^6} L' \sqrt{1
- \frac{B^2 R^4}{(\rho^2 + L^2)^2}}}{\sqrt{1+L'^2}}\right)
-\frac{2B^2 R^4 L \sqrt{\rho^6-\rhoIR^6} \sqrt{1+L'^2}}{(\rho^2 + L^2)^3 \sqrt{1 -
\frac{B^2 R^4}{(\rho^2 + L^2)^2}}}=0\,.}{eq:embedB0i}
In practice we obtain the Minkowski embeddings by shooting out from the L-axis to infinity, while the singular shell embeddings are obtained by shooting inwards and outwards starting from the shell, with the condition that the embedding crosses the shell perpendicularly\footnote{In fact, as we checked, different crossing angles affect the UV and IR behaviour only minimally.}.

\FIGURE[ht]{
\includegraphics[width=15cm,clip=true,keepaspectratio=true]{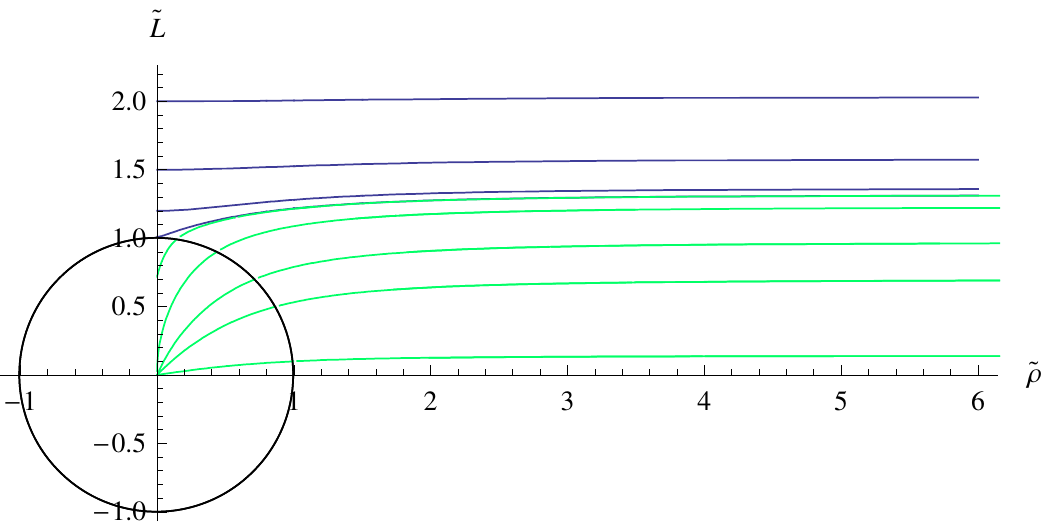}
\caption{D7 brane embeddings for the electric B field at zero temperature. The blue curves are Minkowski embeddings, while the green ones reach the singular shell. The singular shell attracts the D7 brane probes 
as in the finite temperature case, and bends them towards the origin.}
\label{fig:D7flowszero}
}
\FIGURE[ht]{
\includegraphics[width=15cm,clip=true,keepaspectratio=true]{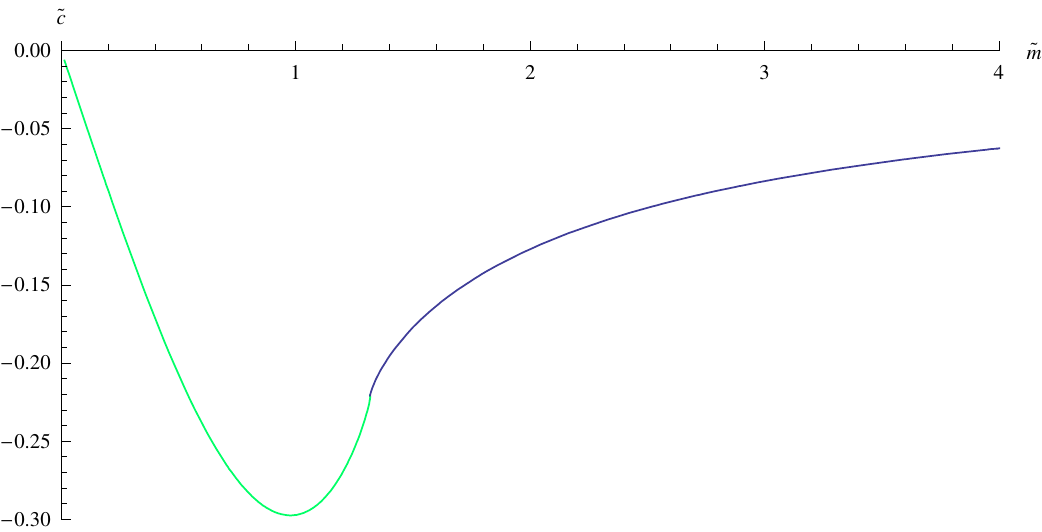}
\caption{$\ctilde(\mtilde)$ for the embeddings in the electric zero temperature case. The
blue line corresponds to the Minkowski solutions, and the green line to those flowing into the shell.}
\label{fig:cvsm0i}
}

\FIGURE[ht]{
\includegraphics[width=15cm,clip=true,keepaspectratio=true]{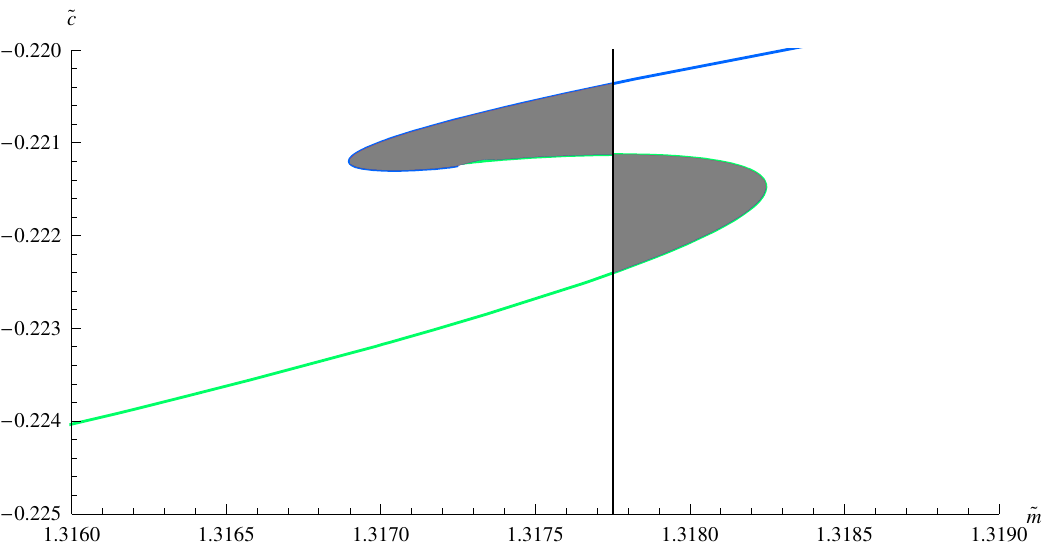}
\caption{{$\ctilde(\mtilde)$ for both physical and nonphysical embeddings in the region $1.317\le\mtilde\le1.322$. 
 The ``S''-shaped bend indicates a first order phase transition at $\mtilde \approx 1.31775$, found by the ``equal area law'' (see text).}}
\label{fig:ptelectricinside}
}

{We extract the UV
asymptotic values of $\mtilde$ for each flow according to \eqref{eq:UVT0}. In figure~\ref{fig:cvsm0i} we plot the condensate $\ctilde$ as a function of
 the reduced quark mass $\mtilde$. The blue curve again corresponds to the Minkowski
 embeddings, while the green curve is for the embeddings flowing into the
 shell of vanishing action. There is a first order phase transition
 at the point where both curves join. This region is shown in detail in
 figure~\ref{fig:ptelectricinside}, where we find a three-fold degeneracy for a
range of masses between about $\mtilde=1.316$ and $\mtilde=1.319$,
in exact analogy to the finite temperature case
\cite{hep-th/0605046}, where $\tilde{m}$ is defined as in (\ref{eq:UVT0}). The exact point of the phase transition can be found by the equal area method:\footnote{{See references \cite{MyersMateos,Johnson1} for the application of the equal area law to such situations.}}
 As $\ctilde(\mtilde)$ in the holographic context is proportional to the first derivative of the free energy, the area below this curve is proportional to the free energy itself. In a region with several possible embeddings, the phase transition then occurs where the difference of the free energies of the two phases has a zero, i.e. where the areas below the $c(m)$ curve are equal. We find
 that below a reduced quark mass of $\mtilde = 1.31775$, the solutions flowing
 into the shell of vanishing action become energetically favoured.}

\FIGURE[ht]{
\includegraphics[height=5cm,clip=true,keepaspectratio=true]{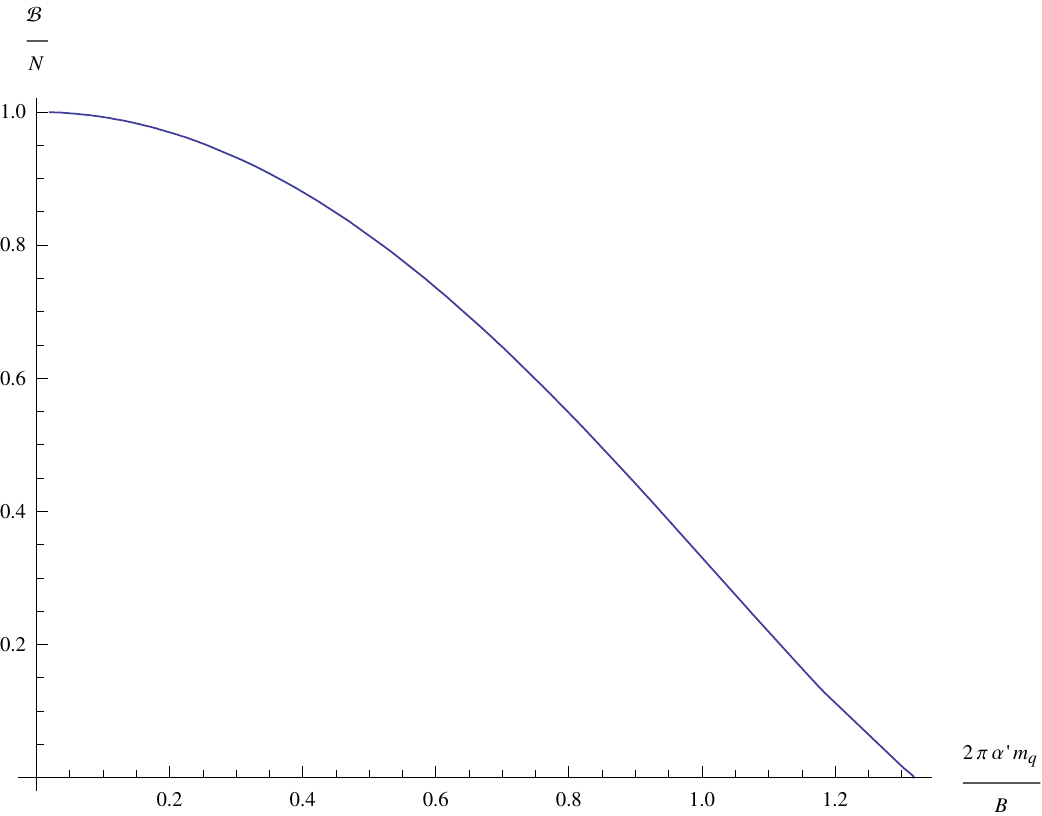}
\caption{Induced baryon number current over reduced quark mass for zero temperature and baryon density.}
\label{fig:inducedcurrent}
}
The induced baryon number current $\langle J_x \rangle = \cB$, which is shown in figure~\ref{fig:inducedcurrent}, shows an interesting behaviour too: For a fixed value of the B field it decreases with growing quark mass until the phase transition point is reached, where the Minkowski embeddings, which have zero induced current, smoothly take over. Note that this plot also includes the singular shell embeddings, whose physical fate still needs to be decided, cf.~section~\ref{sec:conc}. In any case, a kind of threshold for the induced current at zero temperature and baryon density is expected, as vacuum pair production should be impossible for high enough quark masses.

\subsubsection{Meson spectrum}

We now turn to the calculation of the meson spectrum in the presence of the 
electric Kalb-Ramond field at zero temperature.
Up to small modifications, the calculation is very
similar to the one performed for the magnetic case in \cite{hep-th/0701001}.
For the fluctuations we use the ansatz
\begin{eqnarray}
L(\rho)&=&L_0(\rho) \, ,\nonumber\\
\Phi&=&\phi = h(\rho)e^{ik_0t-ik_1x_1}Y_l(S^3)\, , \label{eq:fluc}
\end{eqnarray}
{for which we show the decoupling from the $L$- and gauge field fluctuations on Minkowski embeddings in appendix~\ref{app:B}.}
Here $L_0$ is the embedding obtained from  the equation of motion
(\ref{eq:embedB0i}). Note that in (\ref{eq:fluc}) we now consider a
meson in the $(t,x_1)$ plane, as opposed to the magnetic case
(\ref{eq:magfluc}) where the meson is in the $(x_2,x_3)$ plane.
Linearising the D7-brane action in an analogous way to the
calculation performed in \cite{hep-th/0701001}, we obtain the
equation of motion for the excitations $\phi$,
\begin{equation}\label{eq:eigeneq}
\frac{1}{g}\partial_\rho\left(\frac{gL_0^2\partial_\rho
h(\rho)}{1+{L'_0}^2}\right)-\frac{L_0^2l(l+2)}{\rho^2}h(\rho)+\frac{R^4L_0^2M_{01}^2}{\left(\rho^2+L_0^2\right)^2-R^4B^2}h(\rho)=0
\, .
\end{equation}
Here  $M_{01}^2=k_0^2-k_1^2$  is the meson mass and
$g=\rho^3\sqrt{1+{L'_0}^2}\sqrt{1-\frac{B^2R^4}{\left(\rho^2+L_0^2\right)^2}}$.
We consider only s-wave fluctuations, corresponding to
pseudoscalar mesons ($l=0$), and calculate the spectrum for Minkowski embeddings. The results for $BR^2=1$ are displayed in
Figure \ref{fig:spectrum0i}. 
The dashed lines  in figure \ref{fig:spectrum0i}
show the analytic AdS solutions, $\Mtilde = 2
\mtilde\sqrt{(n+1)(n+2)}$.
\FIGURE[ht]{
\includegraphics[width=12cm,clip=true,keepaspectratio=true]{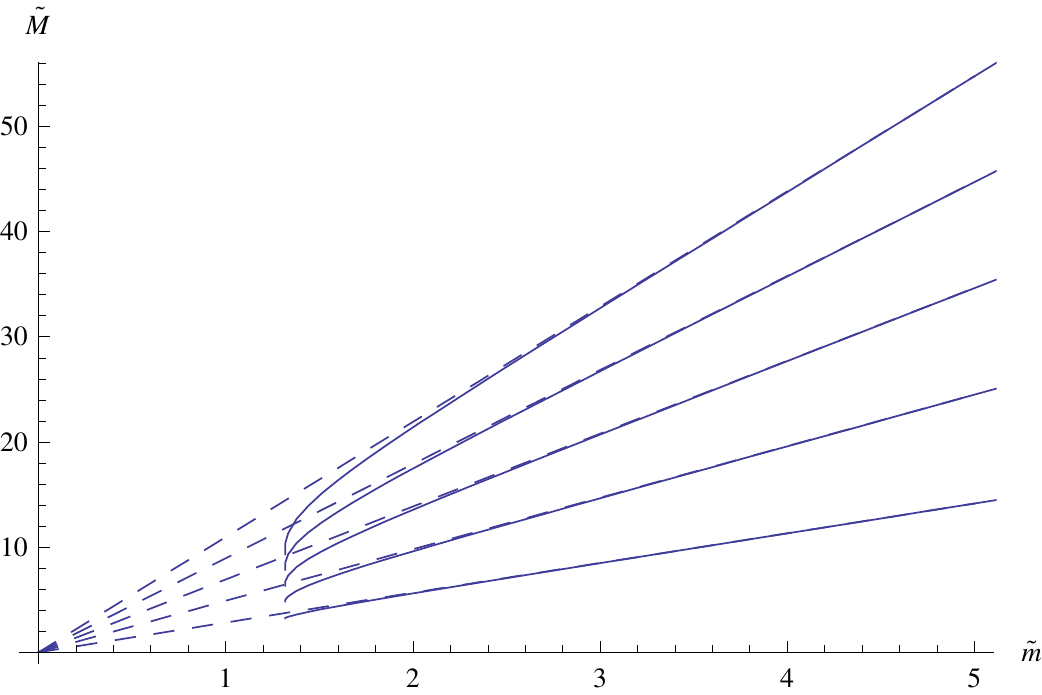}
\caption{First five meson states in the electric ansatz background for the Minkowski embeddings.
$\tilde{M}=\frac{R}{\sqrt{B}}M_{01}$, $\tilde{m}=\frac{m_q
2\pi\alpha'}{\sqrt{B}R}$.} \label{fig:spectrum0i}
}
The behaviour shown in figure \ref{fig:spectrum0i} is consistent
with intuitive expectations: For large reduced quark mass
$\tilde{m}$ the D-branes do not feel the forces exerted by the
singular shell (see figure \ref{fig:D7flowszero}) and thus are
approximately flat, as dictated by supersymmetry of the pure AdS
setting \cite{hep-th/0205236}. 

Note that in contrast to the effect of a magnetic Kalb-Ramond field, the mesons in the presence of the electric field are lighter than without applied field. The electric field thus reduces the binding energy. 
For the embeddings ending inside the shell, we expect the mesons to destabilize through
a mechanism similar to ionization, in analogy to the black hole scenario. 
We leave this for further study. 

\subsection{Stark effect}\label{sec:StarkEffect}

For  weak external electric fields at zero temperature, we now analytically 
calculate the meson mass shift of the $n=l=0$ state, which in the corresponding region of large mass is the
 lightest meson in figure \ref{fig:spectrum0i}, and thus the ground state. 
For this purpose, we use a technique similar to first order perturbation theory familiar from
quantum mechanics, which in this context was suggested to us by Derek Teaney
\cite{Teaney}.

We start by looking at the perturbation exerted by the electric field on a
Minkowski embedding, i.e. an embedding that has high enough quark mass such
that it does not reach the singular shell. Its fluctuation  spectrum is then
discrete, such that the meson masses are well-defined.
To expand the equation for the $\Phi$ fluctuations \eqref{eq:eigeneq} to
lowest non-trivial
order in the electric field, which in fact is ${\cal O} (B^2)$,
we use the following perturbative solution for the brane embedding to
$\cO(B^2)$,
\eq{L = m - \frac{B^2R^4}{4m(\rho^2+m^2)}\,.}{eq:D7embed0iOrderB2}
This result is easily obtained by expanding the embedding equation \eqref{eq:embedB0i} up to second order in $B$. The $\cO(B)$ term would not satisfy the Minkowski embedding boundary condition $L'(0)=0$ and thus has to vanish.
By comparing \eqref{eq:D7embed0iOrderB2} with \eqref{eq:UVas} we find that
 the condensate at small field strength or equivalently large quark mass is
\eq{c(m) = - \frac{B^2R^4}{4m}\,.}{eq:0iT0csmallB}
\FIGURE[ht]{
\includegraphics[width=12cm,clip=true,keepaspectratio=true]{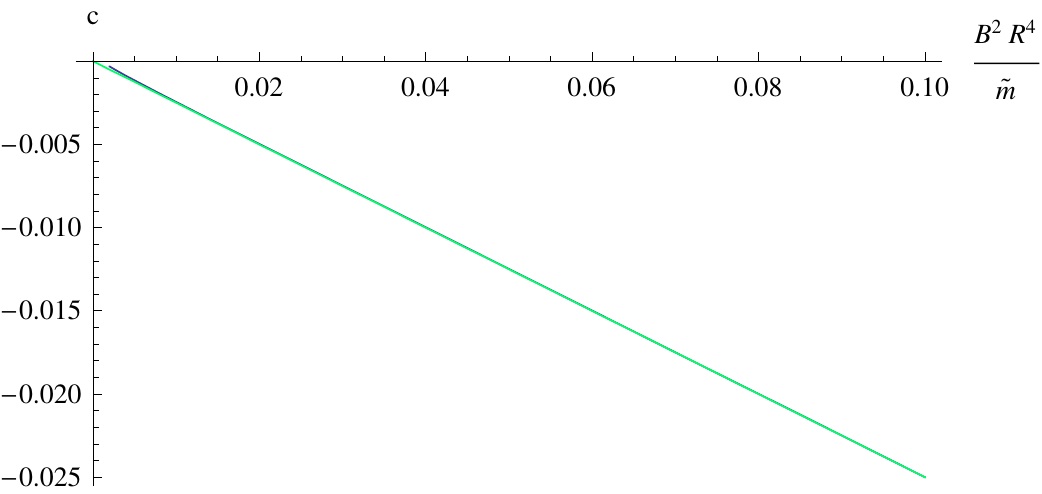}
\caption{The condensate as a function of $\frac{B^2R^4}{m}$, both numerically (dark curve) and in the weak field approximation  \eqref{eq:0iT0csmallB} (green curve). Both curves have a gradient of -4.}
\label{fig:cvsmweakB}
}
Figure \ref{fig:cvsmweakB} shows the quark condensate as a function
of $\frac{B^2R^4}{m}$. The dark curve corresponds to numerical data,
while the light green curve is the weak field result
\eqref{eq:0iT0csmallB}. Both curves coincide to high precision,
and both have a slope of minus four, thus validating
\eqref{eq:0iT0csmallB}. The slight mismatch
near the origin, i.e. in the small $B$ or large $m$ region where the
approximation should hold very well, is due to numerical
instabilities.

Using \eqref{eq:D7embed0iOrderB2} as well as the identities
\eqref{eq:IdStark1}-\eqref{eq:IdStark3} in appendix A,
the equation for the $\Phi$-fluctuations \eqref{eq:eigeneq} up to second order in $B$ reads
\begin{multline}\label{eq:eigeneqB2}
\rho^{-3} \partial_\rho \left( \rho^3 h'(\rho)\right) + \frac{(M_0^2 + \delta(M^2))R^4 h(\rho)}{(\rho^2+m^2)^2} = \\ \frac{B^2R^4}{2m^2}\left[- \frac{M_0^2 R^4 (3m^2-\rho^2) h(\rho)}{(\rho^2+m^2)^4}
- \frac{4 \rho m^2 h'(\rho)}{(\rho^2+m^2)^3} \right.
\left. + \rho^{-3}\partial_\rho\left( \frac{\rho^3 h'(\rho)}{\rho^2+m^2}\right)\right]\,.
\end{multline}
Here we split the exact mass of the ground state $M^2 = M_0^2 +
\delta(M^2)$ into the unperturbed AdS piece $M_0^2 = 8 m^2/R^4$ and
a lowest order correction. It is convenient to introduce
dimensionless quantities\footnote{Note $m = m_q/(2\pi\alpha')$ has
dimension length.} $\rhotilde = \rho/m$, $\Mtilde^2 = R^4 M^2/m^2$
and $\Btilde^2 = B^2R^4/(2m^4)$. In these units and after
multiplying by $\rhotilde^3 m^2$, eq.~\eqref{eq:eigeneqB2} becomes
\begin{multline}
\partial_{\rhotilde} \left( \rhotilde^3 h'(\rhotilde) \right) + W(\rhotilde)(\Mtilde_0^2 + \delta(\Mtilde^2))  h(\rhotilde) = \\
\Btilde^2 \left[- \frac{\Mtilde_0^2 W(\rhotilde) (3-\rhotilde^2) h(\rhotilde)}{(\rhotilde^2+1)^2}
- \frac{4 \rhotilde W(\rhotilde) h'(\rhotilde)}{\rhotilde^2+1}
 + \partial_{\rhotilde}\left( W(\rhotilde) h'(\rhotilde)(\rhotilde^2+1)\right)\right]\,,\label{eq:fluctdimless}
\end{multline}
where we defined the weight function $W(\rhotilde) =
 \frac{\rhotilde^3}{(1+\rhotilde^2)^2}$. We immediately see that the lowest
 order shift in masses will be
 proportional to $\Btilde^2$. We now use a strategy similar to first order
 perturbation theory in quantum mechanics: The fluctuation equation for the pure $AdS_5\times S^5$ case
$$\partial_{\rhotilde} \left( \rhotilde^3 h'(\rhotilde) \right) +
W(\rhotilde)\Mtilde_n^2   h(\rhotilde) = 0 \, , $$
is a Sturm-Liouville problem.  This implies \cite{ArfkenWeber} that its normalizable eigenfunctions
\eq{h_n(\rhotilde) = c_n (1+\rhotilde^2)^{-(n+1)}
{}_2F_1(-(n+1),-n,2,-\rhotilde^2)\, , \Mtilde_n^2 =
4(n+1)(n+2)\,,}{eq:AdSfluctsol}
which have been found in \cite{hep-th/0304032}, can be used to define an orthonormal basis of functions $f_n(\rhotilde)$ on the interval $\rhotilde \in [0,\infty)$ w.r.t to the inner product
\eq{(f,g) = \int\limits_0^\infty \extd\rhotilde \, W(\rhotilde) f(\rhotilde) g(\rhotilde)\,.}{eq:SturmInnerProduct}
Here ${}_2F_1(a,b,c,z)$ is the Gauss hypergeometric function.
Let $\{f_n, n=0,1,...\}$ be such a set, satisfying orthonormality 
\eq{(f_n,f_m) = \delta_{nm}\,,}{eq:fnorthonormality}
 and let us normalize the ground state wave function $h_0(\rhotilde)$ by choosing the coefficient $c_0 = \sqrt{12}$ such that $f_0 = h_0$.
In what follows we will only need the explicit form of this ground state wave function\footnote{Note that ${}_2F_1(-1,0,2,-\rhotilde^2) = 1$.}
\eq{f_0(\rhotilde) = \frac{\sqrt{12}}{1+\rhotilde^2}\,,}{eq:f0}
and the fact that the above orthonormal basis exists \cite{ArfkenWeber}. The
idea is now that the normalized ground state wave function $f_0$ only gets
perturbed by a small amount, which is encoded in the ansatz $h(\rhotilde) =
  a_0 f_0 + \sum\limits_{n>0} a_n f_n{\rho}$, where $a_0 = 1 + \cO(\Btilde^4)$
  and $a_{n} = \cO(\Btilde^2), n>0$. Plugging this ansatz into
 \eqref{eq:fluctdimless}, keeping only terms $\cO(\Btilde^2)$ and using that the $f_n$ satisfy the equation\footnote{Note that ${\Mtilde}'^2_n \neq {\Mtilde}^2_n$ in general, but $\Mtilde'^2_0 = {\Mtilde}_0^2$ because of our choice $f_0=h_0$.}
$$\partial_{\rhotilde}\left(\rhotilde^3 f_n\right) = - \Mtilde'^2_n f_n W\,,$$
we find
\begin{multline}
\sum\limits_{n>0} a_n f_n (\Mtilde'^2_n - \Mtilde^2_0) W  + \delta(\Mtilde^2) W f_0 =\\
 \Btilde^2 \left[ -\frac{\Mtilde_0^2 W f_0 (3-\rhotilde^2)}{(1+\rhotilde^2)^2} - \frac{4 W \rhotilde f_0'}{1+\rhotilde^2} + \partial_{\rhotilde}\left( W(\rhotilde) f_0'(\rhotilde)(\rhotilde^2+1)\right)\right]\,.
\end{multline}
Now multiplying with $f_0$ and integrating over $\rhotilde$ yields, after using orthonormality relation \eqref{eq:fnorthonormality}, an expression for the mass shift
\begin{gather}
\delta(\Mtilde^2) = \Btilde^2 \Big[\underbrace{-\Mtilde_0^2 \left( f_0, f_0 \frac{3-\rhotilde^2}{(1+\rhotilde^2)^2}\right)}_{I_1}
 \underbrace{ -4 \left( f_0, \frac{\rhotilde f_0'}{1+\rhotilde^2} \right) }_{I_2}
  \nonumber\\
\hspace{4.5cm}  + \underbrace{\left(\frac{f_0}{W},\partial_\rho\left( W(\rhotilde)
 f_0'(\rhotilde)(\rhotilde^2+1)\right) \right) }_{I_3}\Big]\,.
\end{gather}
The individual contributions are ($\Mtilde_0^2 = 8$)
\beqas
I_1 &=& - 96 \int\limits_0^\infty \extd \rho
\frac{\rho^3(3-\rho^2)}{(1+\rho^2)^6} = - \frac{28}{5} \, , \\
I_2 &=& - 48  \int\limits _0^\infty \extd\rho
\frac{\rho^4}{(1+\rho^2)^4}\left(\frac{1}{(1+\rho^2)} \right)' = 96
\int\limits_0^\infty \extd \rho \frac{\rho^5}{(1+\rho^2)^6} = \frac{8}{5} \, ,
\\
I_3 &=& -24 \int\limits _0^\infty \frac{\extd\rho}{(1+\rho^2)} \left(\frac{\rho^4}{(1+\rho^2)^3} \right)' = -48 \int\limits _0^\infty \extd\rho \frac{\rho^3 ( 2 - \rho^2) }{(1+\rho^2)^5} = - 2\,.
\eeqas
In dimensionless units, the mass of the ground state is thus shifted by
\eq{\delta(\Mtilde^2) = -6 \Btilde^2  = -3 \frac{B^2R^4}{m^4} \,.}{eq:Mshiftdimless}
Reinstating physical units yields
\eq{\delta M = - \frac{3}{4\sqrt{2}}\frac{B^2R^2}{m^3} = - \frac{3}{16\sqrt{2}\pi^{5/2}}\frac{B^2\sqrt{\lambda}}{\alpha'^2m_q^3} \approx - 0.00758 \frac{B^2\sqrt{\lambda}}{\alpha'^2m_q^3} \,.}{eq:Mshift}
There are several points to mention for this result: First, it has the
expected $B^2$ behaviour of the second order Stark effect. Second, it has the
correct sign, i.e. the mass of the ground state is lowered compared to the
pure AdS case (c.f. figure \ref{fig:spectrum0i}). Furthermore, by introducing
physical units for the B field $\bar{B} = B/ 2\pi\alpha'$, which is of
dimension mass squared, the result becomes independent of $\alpha'$ and is
thus finite in the $\alpha'\rightarrow 0$ limit,
\begin{equation}
\delta M \approx - 0.00758 \frac{\bar{B}^2}{m_q{}^3} \, 4 \pi^2 \sqrt{\lambda}
\, .
\end{equation}
\TABLE[hr]{
        \begin{tabular}{|c|c|c|}\hline
            Level $n$ & $c_n$ & $\alpha_n$ \\\hline
            0 & 0.54 & 3.0 \\
            1 & 1.80 & 3.1 \\
            2 & 3.62 & 3.2 \\
            3 & 7.85 & 3.4 \\
            4 & 8.85 & 3.4 \\\hline
        \end{tabular}
    \caption{Fitted coefficients of the Stark shift, where $\delta{M_n}=-\frac{c_nB^2R^2}{m^{\alpha_n}}$.}
    \label{tab:Starkshift}
}
 The dependence on the
't~Hooft coupling $\lambda$ is also easily understood: As the meson masses themselves are
 proportional to
 $1/\sqrt{\lambda}$ \cite{hep-th/0304032} (c.f.~e.g.~\eqref{eq:AdSfluctsol} and note $R^2 = \sqrt{4\pi\lambda} \alpha'$), the
 combination $\bar{B}^2/m_q^3$ must scale like $\lambda^{-1}$, i.e. must be small in the 't Hooft limit $N\rightarrow\infty$ with $\lambda = g_s N = \mathrm{const.}\gg 1$. This means just that the small $\Btilde$ expansion is valid if either the physical B field $\bar{B}$ is small or the quark mass $m_q$ is large.

Let us now compare this analytical
calculation of the Stark effect with the numerical
data displayed in figure  \ref{fig:spectrum0i}.
In the limit of large quark mass $\tilde{m}$, we study the
difference between the AdS meson spectrum and the value of $\tilde{M}$.
Numerically we find that \footnote{Note that in the units of figure
  \ref{fig:spectrum0i}, $B$ is scaled to $B=1$.}
\begin{equation}
\delta \Mtilde = \tilde{M}-\tilde{M}_{AdS}=-\frac{0.54R^2}{
\tilde{m}^3} \, .
\end{equation}
Performing the appropriate rescaling of $\tilde{M}$ and $\tilde{m}$ to
reintroduce B we find that
\begin{equation}
\delta{M}=-\frac{0.54B^2R^2}{m^3}\,.
\end{equation}
From equation (\ref{eq:Mshift}) we find
\begin{equation}
\delta{M}=-\frac{0.53B^2R^2}{m^3}\,,
\end{equation}
in very good agreement with the numerical calculation. The discrepancy is
likely to come from both numerical errors and higher order corrections in
the expansion around small $B$ (large $m$).
Table~\ref{tab:Starkshift}
  shows the coefficients $c_n$ and $\alpha_n$ in the mass shift
\begin{equation} \label{tableshift} \delta
  M = - c_n \frac{B^2R^2}{m^{\alpha_n}}
\end{equation}
 as obtained by fitting (\ref{tableshift})
to the numerical result for the meson masses displayed in figure
\ref{fig:spectrum0i},
also for the higher modes $n\ge1$. Keeping in mind that according to our experience, the
numerical errors get larger for higher states (although we can not specify
the error quantitatively), there is a chance that $\alpha_n$ is $n$-independent
with a value of three, as can be expected on dimensional grounds by
requiring the mass shift to be proportional to the dimensionful
Kalb-Ramond field $\bar{B}^2$ and simultaneously independent of $\alpha'$.

\subsection{Electric Kalb-Ramond field at finite temperature}\label{sec:electricfiniteT}
{
We conclude this section by commenting on the embeddings in the finite temperature
AdS-Schwarzschild background with electric field, which we obtain numerically from the
 action \eqref{eq:SD7Legendre2T}. Since we are introducing baryon number density through the D7 brane gauge field, we have to distinguish between embeddings for zero and finite density, as smooth Minkowski embeddings are only consistent for vanishing density \cite{MyersI}. Note that considering finite densities is the more general case: For ensuring a regular DBI action at the singular shell, $D=0$ would suffice to satisfy equation~\eqref{eq:BDrelationT}.

Figure~\ref{fig:finiteTDzero} shows embeddings calculated for different strengths of the B-field, and vanishing baryon density $D=0$. Since at zero baryon density both Minkowski and singular shell embeddings are consistent (cf. figure~\ref{fig:D7flowszero}), a well-defined zero temperature limit in the canonical ensemble  reproducing figure  \ref{fig:D7flowszero} is possible for $D=0$ only (see the discussion in section~\ref{sec:conc}). As can be seen from figure~\ref{fig:finiteTDzero}, in contrast to finite baryon number density, the black hole embeddings cover only a finite range of asymptotic quark mass up to a maximal value. Above this mass, 
which depends on the B field through $\mtilde = (1.27 \pm 0.03) \sqrt{BR^2+\sqrt{4b^4+B^2R^4}}/\sqrt{2}$, 
Minkowski embeddings take over. This now allows for a well-defined zero temperature limit. Lowering the temperature, the black hole horizon shrinks to the extremal AdS horizon at $\rho=L=0$, where the black hole embeddings then end. The corresponding chiral condensate, in particular its dependence on the field strength, on the temperature and on the quark mass, was calculated in \cite{Johnson2}.
\FIGURE[ht]{
\includegraphics[trim = 7mm 0mm 0mm 0mm, width=\textwidth,clip=true,keepaspectratio=true]{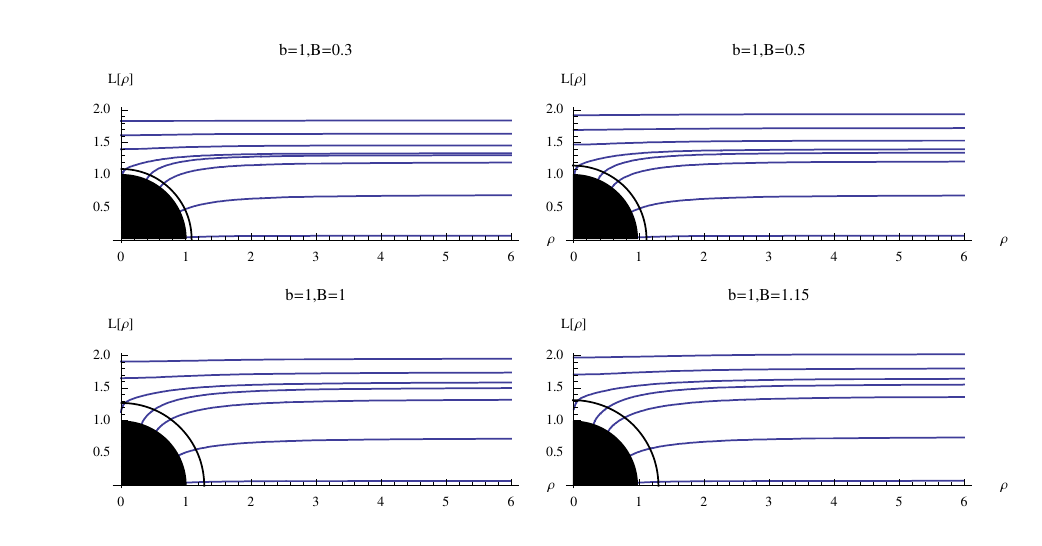}
\caption{{Minkowski and singular shell embeddings for the electric case at finite temperature and zero baryon number density $D=0$.}} \label{fig:finiteTDzero}
}

Figures~\ref{fig:TBflows} and \ref{fig:TBflows20} show black hole embeddings calculated for a baryon number density of $D=10$ and $D=20$, respectively. For comparison we also included Minkowski embeddings which, as we know very well, can only coexist with black hole embeddings  at finite baryon number density in a grand canonical ensemble \cite{MyersII}, where the baryon chemical potential is fixed instead of the number density (cf. also section~\ref{sec:conc}). The black hole embeddings pass through the shell
of vanishing action smoothly and reach the black hole horizon. We observe that black hole embeddings at finite baryon number density can cover the whole range of asymptotic quark masses, a fact already noted without external electric fields in \cite{MyersI}.
Comparing the two figures \ref{fig:TBflows} and \ref{fig:TBflows20}, we find the effect of changing the baryon number density $D$ as given by  \eqref{eq:conservedD}: In both figures, the Minkowski and black hole embeddings were calculated for the same values of the infrared boundary condition $L(\rhoIR) = \LIR$. The green black hole embeddings are sensitive to the value of $D$: For a given infrared  boundary condition, they reach larger asymptotic quark masses $L(\infty)$ for larger $D$. Thus the scaling of $D$ corresponds to a dilation of the energy scale. Note that this scaling effect is also present for black hole embeddings in the canonical ensemble at zero external electric field, c.f.~e.g.~figure~4 in \cite{MyersI}.
\FIGURE[ht]{
\includegraphics[height=10cm,clip=true,keepaspectratio=true]{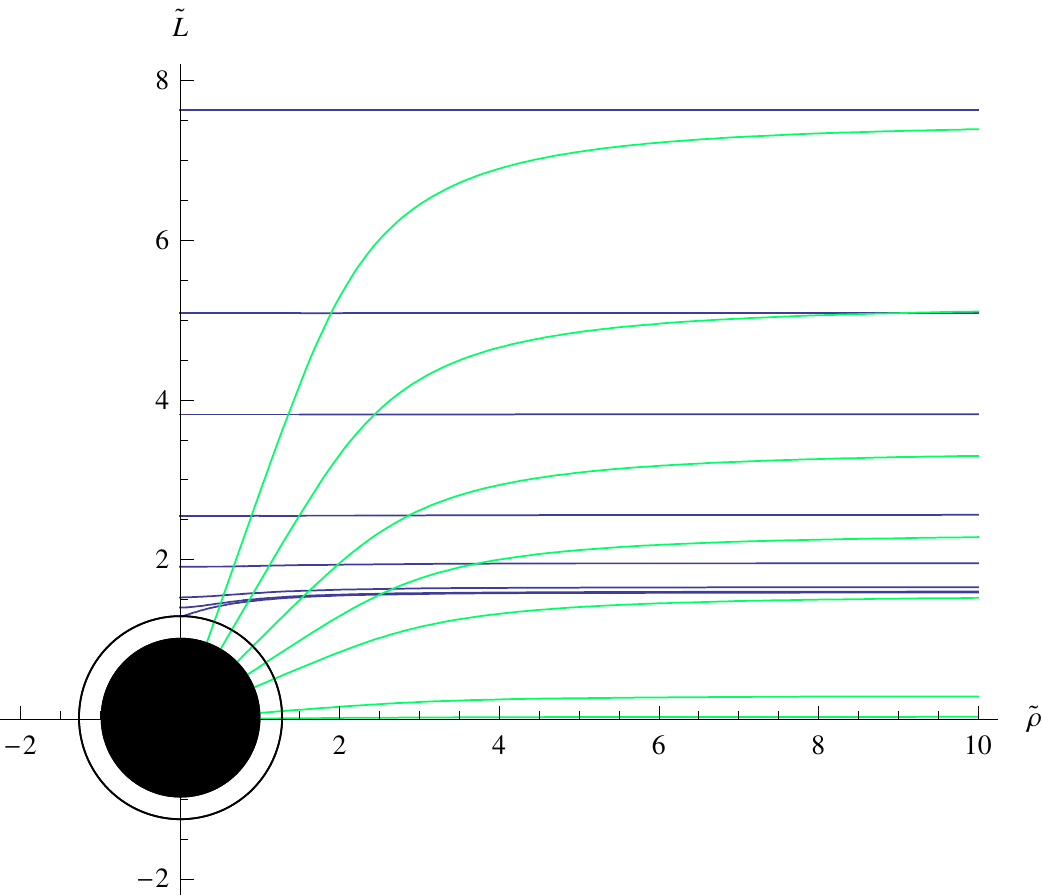}
\caption{{Minkowski ($b=B=R=1$, $D=0$) and black hole embeddings for the electric case at finite temperature ($b=B=R=1$, $D=10$). Note that both types embeddings can only exist simultaneously in the grand canonical ensemble \cite{MyersII}.}} \label{fig:TBflows}
}
\FIGURE[ht]{
\includegraphics[height=10cm,clip=true,keepaspectratio=true]{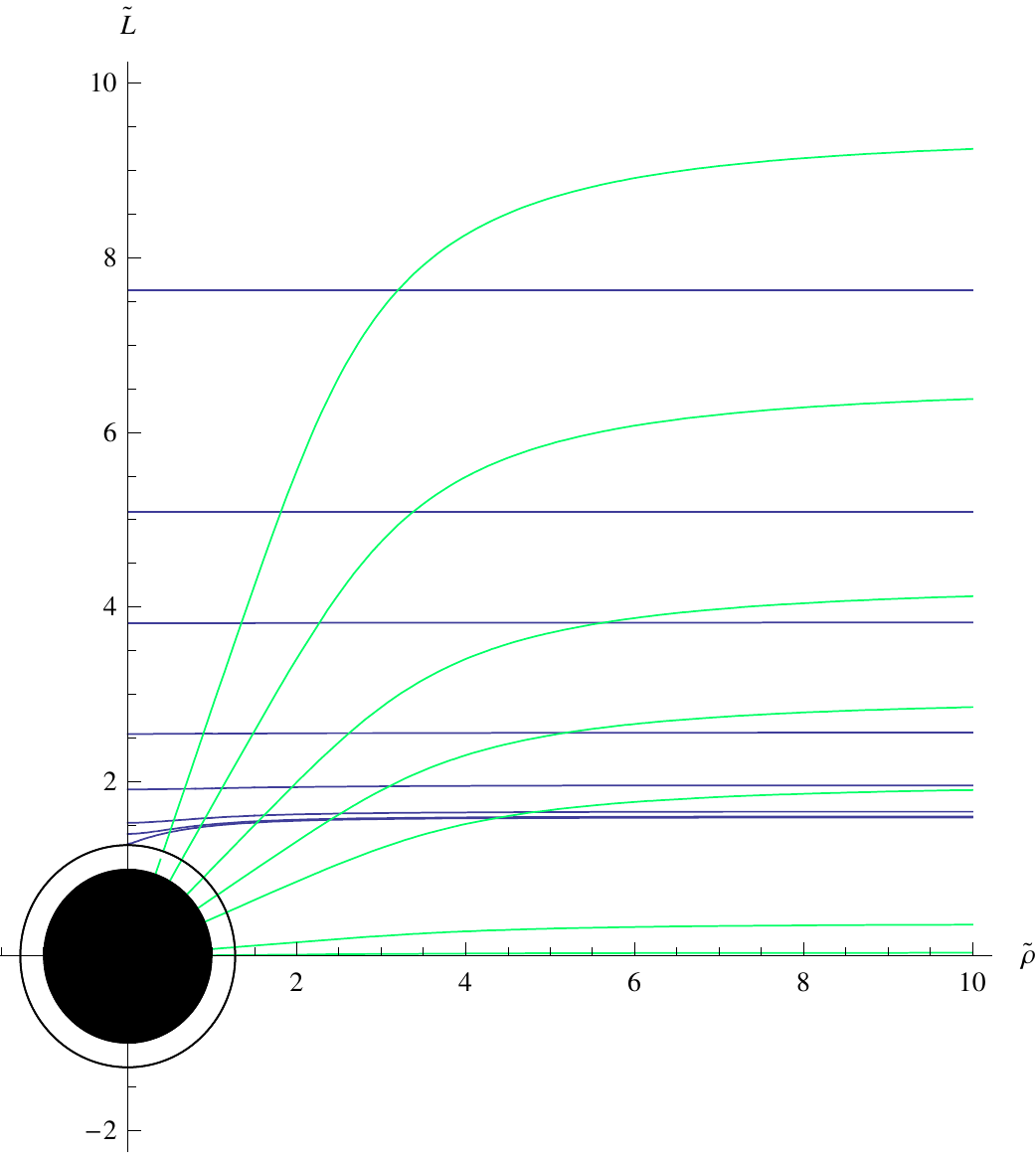}
\caption{{Minkowski ($b=B=R=1$, $D=0$) and black hole embeddings for the electric case at finite temperature ($b=B=R=1$, $D=20$). Note that both types embeddings can only exist simultaneously in the grand canonical ensemble \cite{MyersII}.}} \label{fig:TBflows20}
}
}


\section{Conclusions and Outlook}\label{sec:conc}

The magnetic (spatial-spatial) and electric (temporal-spatial) external B
fields in the directions parallel to the boundary have quite different effects
on D7 probe brane embeddings. In the magnetic case there is a
repulsion, such that even at finite temperature spontaneous
chiral symmetry breaking occurs. The magnetic field has a confining effect on
the mesons.

In the electric case there is a singular
shell on which the DBI action vanishes. This shell has an attractive effect on
the embeddings, which may be interpreted as ionization.
A gauge field has to be switched on to ensure a regular action inside the
shell. The {full} stability analysis of this scenario is delicate, 
as strings in electric fields are known to display instabilities,
and we leave this analysis for further study here. 

{Let us, however, comment on the question of the thermodynamical ensemble considered in the electric case. As shown in \cite{MyersI}, switching on a gauge field $A_t$ corresponds to including baryon number density and a baryon chemical potential into the thermodynamics of the dual gauge theory. The canonical ensemble is then characterized by a fixed number density and a varying chemical potential, while in the grand canonical ensemble the chemical potential is held fixed \cite{MyersNeu}. Furthermore, Minkowski embeddings which close off smoothly at $\rho=0$ are only consistent for vanishing baryon number density: A nonvanishing baryon number density would correspond to strings creating a cusp at $\rho=0$ on the brane while stretching down to the black hole  horizon \cite{MyersI}, or to the AdS horizon in the zero temperature limit. Thus, as we interpret our results with fixed number density $D$ as in \cite{KarchOBannon}, we are naturally considering the system  in the canonical ensemble.

For vanishing baryon density $D=0$, when Minkowski embeddings are consistent, the embeddings which reach the singular shell would   necessarily have vanishing baryon number density too. However, due to eq.~\eqref{eq:BDrelationT}, these embeddings correspond to a nonzero induced baryon number current $\langle J_x \rangle=\cB$. This describes the situation after meson melting, where the quarks and antiquarks are accelerated by the applied external Kalb-Ramond field. This is equivalent to vacuum pair production of baryon number charge by thermal effects and by the applied external field. This interpretation coincides with the result of \cite{KarchOBannon} that the baryon number conductivity is finite even in the $n_q\rightarrow 0$, $T\rightarrow 0$ limit.

At zero baryon number density,  figures~\ref{fig:D7flowszero}~and~\ref{fig:finiteTDzero} show that above a certain asymptotic quark mass, the embeddings do not reach the singular shell any more.  We then expect the dissociation phase transition to happen between a singular shell embedding and a Minkowski one. This again coincides with the necessity to ascribe a vanishing baryon number density to the singular shell embeddings in the canonical ensemble, as the dissociation of a meson, i.e. a quark-antiquark pair, will not create finite quark number densities\footnote{The quark and baryon number densities are related by a factor of $n_q=N_c n_b$.}. In the zero temperature limit,  the black hole embeddings end at the point $L=\rho=0$ (see figure~\ref{fig:D7flowszero}). This is where the extremal black hole, i.e. the D3-brane horizon sits.

As mentioned before, for finite temperature and nonvanishing baryon density there exist only black hole embeddings  in the canonical ensemble (cf. figure~\ref{fig:TBflows} and \ref{fig:TBflows20}). In this case, we expect the dissociation phase transition to occur between two black hole embeddings at least for sufficiently small electric field, where the singular shell is near the horizon and the physics are continuously related to the zero-field case \cite{MyersI}. 

From the above, we therefore think that an interesting future task is to explore the phase diagram for finite external electric field in the grand canonical ensemble. We expected that Minkowski embeddings might play a role for the dissociation transition at finite temperature similar to the situation described in \cite{MyersNeu}, and that the dissociation of the mesons at high temperatures, or equivalently at small quark masses, could lead to nonzero baryon number densities due to the fixed chemical potential. Note, however, that for finite electric field the application of equilibrium thermodynamics is justified only in the large $N_c$-limit with $N_f/N_c\rightarrow 0$. In this case, the $\cN=4$ plasma can absorb an infinite amount of energy from the $\cN=2$ baryon number charge carriers (see the conclusions of \cite{KarchOBannon} for details).

The results for the electric Kalb-Ramond field obtained in this paper, come, however, with an open question which should be answered in future work: It remains to analyse whether the conically singular embeddings, whose presence was noticed first in \cite{Johnson2}, and are also seen in our results, most easily in  figure~\ref{fig:D7flowszero}, are physical or not. These embeddings occur in the intermediate quark mass range between black hole and Minkowski embeddings. In order to be physical, for instance describing a new intermediary phase of melting mesons in the presence of an external $U(1)$ field, they have to be in a force-balanced state of the D7 brane with strings stretching from the cusp to the horizon. For vanishing B field this situation was analyzed in \cite{MyersI} (around eq.~(2.40) there), with the result that such static configurations do not exist. The calculation in \cite{MyersI} relies on the fact that, in the region $\rho\approx 0$, the Legendre-transformed D7-brane action reduces to an action for a density $n_q$ of Nambu-Goto strings smeared in the Minkowski directions and stretching along the L-axis from the black hole horizon to the cusp. In our case, in the same limit, eq.~\eqref{eq:SD7Legendre2T} becomes\footnote{Note the change of coordinates $L=r \cos \theta$, $\rho=r \sin \theta$.}
\eq{\bar{L}'_{D7} \simeq D\sqrt{\left(g_{tt}(r)-g_{tt}(\rIR)\right)\left(1 - \frac{B^2}{g_{tt}(r)g_{xx}(r)}\right)\left(g_{\rho\rho}(r)+g_{\theta\theta}(r)(\partial_r \theta)^2 \right)}\,.}{eq:LimitD7toF1}
Without the $B^2$-term in the square root, this would reduce to a Nambu-Goto type action for strings smeared in the Minkowski directions with density D. It might, however, be possible that the strings get dragged along in the x-direction by the B field, and thus a  trailing string dragged through the plasma could be attached at the cusp. Also stringy $\alpha'$-corrections to the DBI action \cite{DBIcorrections} might resolve the conical singularity in the embedding.
}

Generically, also from the field theory perspective, an instability is 
expected in the electric case, since turning on an external electric Kalb-Ramond
field also requires the presence of a non-zero gauge field background corresponding to a non-vanishing baryon number density and baryon chemical potential $\mu$. It is known for supersymmetric
 theories, which involve scalars, that this leads to an upside-down potential
 of the form \cite{SUSY}
\begin{equation}
V(\phi) = - \mu^2 |\phi|^2 \, ,
\end{equation}
which will, in the end, lead to Bose-Einstein condensation of the scalars.
For an isospin chemical potential, this instability was reproduced in the
holographic context in \cite{AEEG}. Also in the study of baryon chemical
potentials, thermodynamical instabilities of D7 brane embeddings at
which tachyonic meson modes appear have been
found \cite{MyersI,MyersII}. These latter  instabilities
are of a more intricate structure, since
they are confined to particular regions of the phase diagram.

The detailed structure of the phase diagram for the electric external field is
thus very subtle. It will be interesting to study it further, in particular in the light of the very recent work \cite{KarchSeptember,MyersNeu}. {Some new results in this direction are presented in \cite{Johnson2}.}


\acknowledgments
We are grateful to Tameem Albash, 
Paolo Aschieri, Luzi Bergamin, Jos\'e Edelstein, Nick Evans, Veselin Filev, Johannes Gro{\ss}e, Daniel Grumiller, Clifford
Johnson, Matthias Kaminski, Paul Koerber, Arnab Kundu, Javier Mas, Jeong-Hyuck Park, Felix Rust, 
Derek Teaney and Maxim Zabzine for discussions. J.~E.~is grateful to the
Isaac Newton Institute in Cambridge, England, for hospitality
during part of this project. R.~M.~thanks the Erwin-Schrödinger-Institute in Vienna for its hospitality. 
This work was supported in part by the Cluster of Excellence ``Origin and Structure of the Universe''. 

\appendix

\section{Some useful expansion formulae}\label{app:A}

To expand eq.~\eqref{eq:eigeneq} while calculating the Stark shift, the following identities which hold up to $\cO(B^2)$ are useful:
\beqa\label{eq:IdStark1}
g &=& \rho^3\sqrt{1+L'^2}\sqrt{1 - \frac{B^2R^4}{(\rho^2+L^2)^2}} \nonumber \\
&\simeq& \rho^3 \left( 1 - \frac{B^2R^4}{2(\rho^2+m^2)^2})\right)\,, \\\label{eq:IdStark2}
g^{-1} &\simeq& \rho^{-3} \left( 1 + \frac{B^2R^4}{2(\rho^2+m^2)^2} \right)\,,\\\label{eq:IdStark3}
\left[(\rho^2+L^2)^2 - B^2R^4 \right]^{-1} &\simeq& (\rho^2+m^2)^{-2} \left( 1 + \frac{B^2R^4}{(\rho^2+m^2)^2}\right)\,.
\eeqa

\section{{Decoupling of $\Phi$-fluctuations}}\label{app:B}

{
In this appendix we show that for the magnetic and electric Kalb-Ramond field \eqref{eq:KRAnsatz}, both at finite and zero temperature, the $L$- and gauge field field fluctuations can be decoupled from the $\Phi$-fluctuations and consistently set to zero by the respective ansätze \eqref{eq:magfluc} and \eqref{eq:fluc}. For the electric case, as we will show, this holds only for the Minkowski embeddings, which have both zero baryon number density and current, i.e. for a trivial gauge field background. For a nontrivial gauge field background, like the one which renders the singular shell embeddings consistent, this decoupling will only be possible if the $\Phi$-fluctuations do not depend on the Minkowski coordinates at all, $\phi=\phi(\rho)$. The calculation follows \cite{hep-th/0701001}. 

To show the decoupling, we have to verify that in the part of the action quadratic in the fluctuations no couplings of $\chi$ and $A$ to $\phi$ appear. We take the embedding fluctuations to be $L = L_0(\rho) + \chi$, $\Phi = \phi$, where $L_0(\rho)$ is the embedding of the D7 brane into the appropriate background. For now, let us assume a trivial gauge background $F=0$ on the brane, such that the gauge field $A$ is a pure fluctuation. For a diagonal background metric $G_{MN}(\rho,L)=\text{diag}(-G_{tt}, G_{xx},G_{yy},G_{zz},G_{\rho\rho},G_{\eta\eta},G_{\xi_1\xi_1},G_{\xi_2\xi_2},G_{LL},G_{\Phi\Phi})$ not depending on $\Phi$ and the magnetic ansatz \eqref{eq:KRAnsatz}, the pull-back of metric and B field can be split into an embedding part, a part linear and a part quadratic in the fluctuations:
\beqa
E_{ab} &=& P[G+B]_{ab} = E^{(0)}_{ab} + E^{(1)}_{ab} + E^{(2)}_{ab}\,, \\
E^{(0)}_{ab} &=& (\cG^{-1}+\theta)^{-1}_{ab}\,, \\\nonumber
(\cG^{-1})^{ab} &=& \text{diag}\left(-G_{tt}^{-1}, G_{xx}^{-1},\frac{G_{zz}}{G_{yy}G_{zz} + B^2},\frac{G_{yy}}{G_{yy}G_{zz} + B^2},(G_{\rho\rho}+{L_0'}^2G_LL)^{-1},\right. \\
&& \hspace{10mm}\left. G_{\eta\eta}^{-1},G_{\xi_1\xi_1}^{-1},G_{\xi_2\xi_2}^{-1} \right)\,, \\
\theta^{23} &=& - \theta^{32} = - \frac{B}{G_{yy}G_{zz} + B^2}\,,\quad\text{others zero}\,,\\
E^{(1)}_{ab} &=& E^{(1)}_{Sab} + E^{(1)}_{Aab} \,,\\
E^{(1)}_{Sab} &=& 2 L_0' G_{LL}(L_0) \de_{\left(a\right.}^\rho \partial_{\left. b \right)} \chi + \chi \partial_L G_{ab}(L_0) + {L_0'}^2 \chi \de_a^\rho \de_b^\rho \partial_L G_{LL}(L_0)\,, \label{eq:E1s}\\
E^{(1)}_{Aab} &=& 0\,, \label{eq:E1a}\\
E^{(2)}_{ab} &=& E^{(2)}_{Sab} + E^{(2)}_{Aab}\,,\\
E^{(2)}_{Sab} &=& G_{\Phi\Phi}(\partial_a \phi)(\partial_b \phi) + G_{LL}(\partial_a \chi)(\partial_b \chi) + 2 \de_{\left( a \right.}^\rho (\partial_{\left. b \right)} \chi ) \chi L_0' \partial_L G_{LL}(L_0) \label{eq:E2s}\\
&& + \frac12 \de_a^\rho \de_b^\rho {L_0'}^2 \chi^2 \partial_L^2 G_{LL}(L_0) + \frac12 \chi^2 \partial_L^2 G_{ab}\,,\\
E^{(2)}_{Aab} &=& 0\,.\label{eq:E2a}
\eeqa
Here, we split the inverse of $E^{(0)}$ further into its symmetric (the ``open string metric'') $\cG^{-1}$  and its antisymmetric part (the ``noncommutativity parameter'') $\theta$, as well as the linear and quadratic fluctuation parts. Note that the antisymmetric parts at linear and quadratic order vanish. 

We now pull out the $E^{(0)}$ from the square root in the DBI action and use the usual determinant expansion to obtain the fluctuation part of the DBI action, dropping a  factor $-T_7/g_s$, 
\beqa\label{eq:L2DBI}
\cL^{(2)}_{DBI} &=& \frac12 \sqrt{-\det E^{(0)}}\Big[ \Tr({E^{(0)}}^{-1}{E^{(2)}}) + \frac{1}{4} (\Tr({E^{(0)}}^{-1}E^{(1)}))^2 \\
&&+ \frac{1}{4} (\Tr({E^{(0)}}^{-1}F))^2 + \frac12\Tr({E^{(0)}}^{-1}E^{(1)})\Tr({E^{(0)}}^{-1}F) \\
&&- \frac12 \Tr({E^{(0)}}^{-1}E^{(1)})^2
- \frac12 \Tr({E^{(0)}}^{-1}F)^2 - \Tr({E^{(0)}}^{-1}E^{(1)}{E^{(0)}}^{-1}F) \Big], \\
\sqrt{-|E^{(0)}|} &=& \sqrt{\left(\prod\limits_{M\in\{t,x,\eta,\xi_1,\xi_2\}} G_{MM}\right)(G_{yy}G_{zz} + B^2)(G_{\rho\rho}+{L_0'}^2G_LL)}\,.
\eeqa
Because the antisymmetric parts \eqref{eq:E1a} and \eqref{eq:E2a} vanish and the $\Phi$-fluctuations only show up at second order in \eqref{eq:E2s}, it is seen by checking term by term that \eqref{eq:L2DBI} includes couplings between $F$ and $\chi$ only, but not between $F$ or $\chi$ with $\phi$, and thus $A=\chi=0$ decouples the angular fluctuations. The $\phi$-part of the DBI action thus reads
\beqas
\cL_{DBI,\phi}^{(2)} & = & \frac12 \sqrt{-\det E^{(0)}} G_{\Phi\Phi}(L_0)({\cG}^{-1})^{ab} \partial_a\phi\partial_b\phi \,,
\eeqas
which reduces to \eqref{eq:fluctmagfiniteT} for the AdS-Schwarzschild background. This reasoning, as is easily checked, also holds for the gauge field background \eqref{eq:gaugeansatz}, which is needed for the electric case.

The dangerous part, which may couple gauge field and angular fluctuations, is the Wess-Zumino part of the action to second order in the fluctuations, 
$$\int\limits_{\cM_8} \frac12 P[C_4]\wedge F\wedge F + P[\tilde{C}_4] \wedge P[B] \wedge F\,,$$
which could in principle introduce couplings of F to $\partial_t \phi$, $\partial_x \phi$ and $\partial_{t} \partial_\rho \phi$ or $\partial_{x} \partial_\rho \phi$, as the magnetic dual of the Ramond-Ramond four-form potential includes $\Phi$-fluctuations, $P[\tilde{C}_4] \propto \partial_a \phi$ (see eq.~(45) in \cite{hep-th/0701001}). With the ansatz \eqref{eq:magfluc}, which does not depend on $t$ or $x$, these couplings vanish. Thus $\chi = 0$, $A=0$ and the ansatz \eqref{eq:magfluc} yields \eqref{eq:fluctmagfiniteT}.

For the electric case, the Wess-Zumino action does lead to additional couplings for a nontrivial gauge field background $F_0$. As for the Minkowski embeddings no gauge background is needed for consistency, the ansatz \eqref{eq:gaugeansatz} used to calculate the spectrum at zero temperature is consistent. The new couplings, which spoil the above reasoning for the singular shell embeddings, are again due to the magnetic dual of $C_4$, namely $\int P[\tilde{C}_4]\wedge F_0 \wedge F$. As $P[\tilde{C}_4]$ containts $\partial_a \phi \extd \xi^a$, this induces the following couplings (schematically): $(\partial_y \phi) (F_{zx} + F_{zt}) + (\partial_z \phi) (F_{xy} + F_{ty})$. These couplings vanish only if $\phi(\rho)$ is a function independent of all Minkowski coordinates, i.e. a zero mode in the Minkowski directions, $(\om,k_1,k_2,k_3)=0$. Furthermore imposing $h(\rho)$ to be normalizable singles out a unique solution. Thus for calculating the fluctuation spectrum of singular shell embeddings, in order to prove the expectation that the mesons are in a dissociated state and to clarify the role of the conically singular embeddings for the expected dissociation transition, it will be necessary to solve coupled partial differential equations for the gauge field and $\phi$, most likely with numerical methods generalizing those used in \cite{Johnson2}.
}


\FIGURE[ht]{
\begin{sideways}
\includegraphics[trim = 4mm 0mm 0mm 0mm, width=22cm,clip=true,keepaspectratio=true]{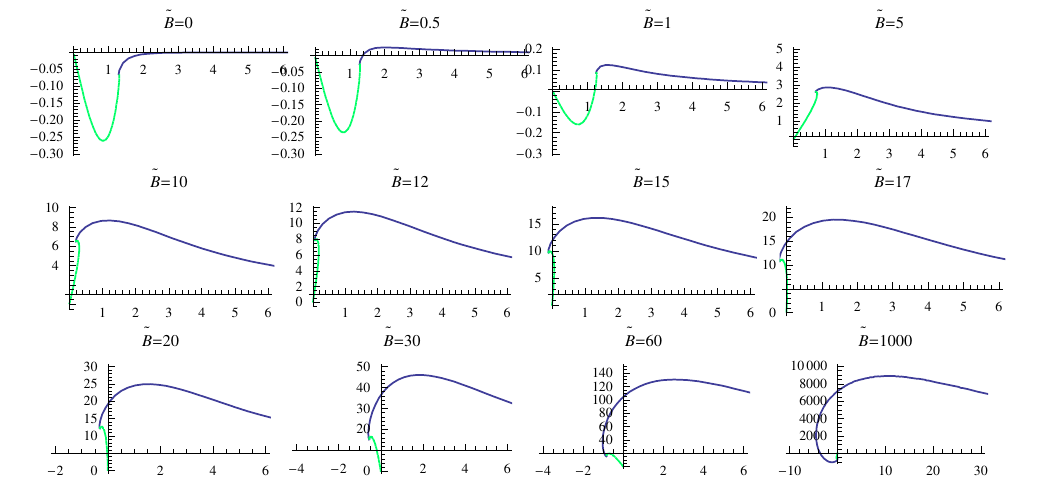}
\end{sideways}
\caption{{Condensate $\frac{8\langle\bar{q}{q}\rangle}{\sqrt{\lambda}N_cT^3}$ against
quark mass $\frac{2m_q}{\sqrt{\lambda}T}$ for the magnetic ansatz
for increasing $\tilde{B}$ from 0 to 1000. There is a critical value
of $\tilde{B}\approx 16 $, above which there is no melted phase. Note the return of the
spiral behaviour seen in figure~\ref{fig:spiral} for large
$\tilde{B}$.}}
\label{fig:cvsmxx}
}

\FIGURE[ht]{
\begin{sideways}
\includegraphics[width=22cm,clip=true,keepaspectratio=true]{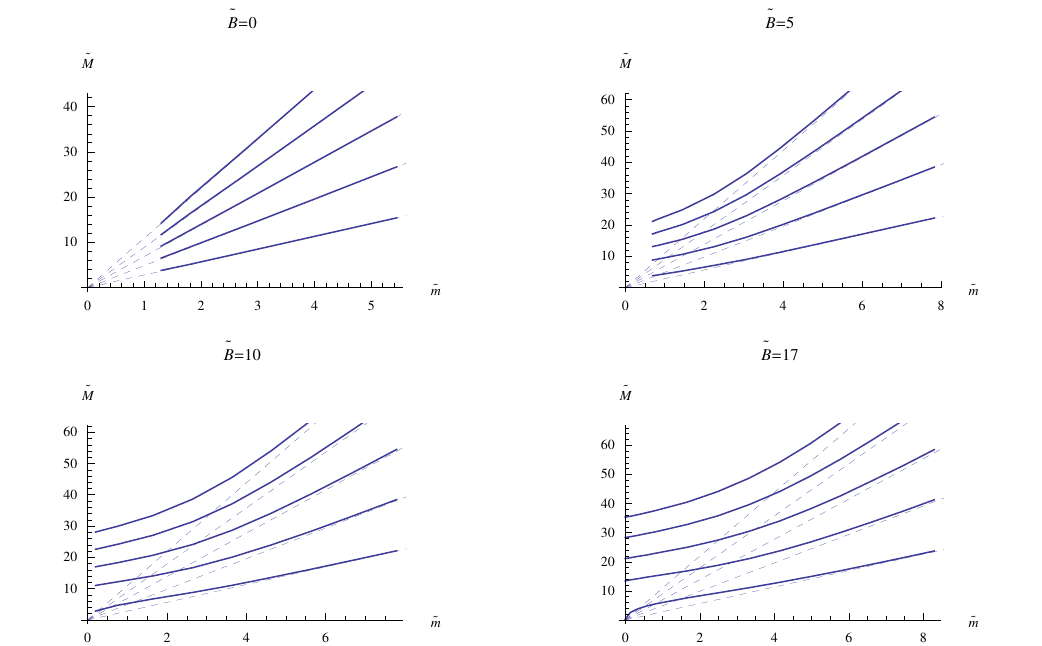}
\end{sideways}
\caption{{Meson masses for the AdS-Schwarzschild black hole background for different magnetic field strengths. The dashed lines are the pure AdS result $\Mtilde = 2
\mtilde\sqrt{(n+1)(n+2)}$, cf.~\cite{hep-th/0304032}. Above the critical field strength $\Btilde_{crit}\approx 16$ the lowest meson state becomes massless at zero quark mass and thus is identified with the Goldstone boson of chiral symmetry breaking.}}
\label{fig:spectra}
}

\end{document}